\documentclass[preprint,english,showpacs,preprintnumbers,amsmath,amssymb,floatfix]{revtex4}
\usepackage[T1]{fontenc}
\usepackage[latin9]{inputenc}
\usepackage{color}
\usepackage{array}
\usepackage{amstext}
\usepackage{graphicx}
\usepackage{esint}

\makeatletter

\@ifundefined{textcolor}{}
{%
 \definecolor{BLACK}{gray}{0}
 \definecolor{WHITE}{gray}{1}
 \definecolor{RED}{rgb}{1,0,0}
 \definecolor{GREEN}{rgb}{0,1,0}
 \definecolor{BLUE}{rgb}{0,0,1}
 \definecolor{CYAN}{cmyk}{1,0,0,0}
 \definecolor{MAGENTA}{cmyk}{0,1,0,0}
 \definecolor{YELLOW}{cmyk}{0,0,1,0}
 }

\@ifundefined{definecolor}
 {\usepackage{color}}{}
\@ifundefined{definecolor}
 {\usepackage{color}}{}
\makeatother

\makeatother

\usepackage{babel}

\begin{document}

\title{Impact of Beauty and Charm H1-ZEUS Combined Measurements on PDFs and Determination of the Strong Coupling}

\author{A.~Vafaee}
\email[]{vafaee.phy@gmail.com}
\affiliation{National Foundation of Elites, P. O. Box 14578-93111, Tehran, Iran}
\author{A.~Khorramian}
\email[]{Khorramiana@semnan.ac.ir}
\affiliation{Faculty of Physics, Semnan University, P. O. Box 35131-19111, Semnan, Iran}

\date{\today}

\begin{abstract}
In this QCD analysis, we investigate the impact of recent measurements of heavy-flavor charm
and beauty cross sections data sets on the simultaneous determination of Parton Distribution Functions
(PDFs) and the strong coupling, $\alpha_s(M^2_Z)$. We perform three different fits based on Variable-Flavour Number Scheme (VFNS) at the Leading Order (LO) and Next-to-Leading Order (NLO) and choose the full HERA run I and II combined data as a new measurement of inclusive Deep Inelastic Scattering (DIS) cross sections for our base data set. We show that including charm and beauty cross sections data reduces the uncertainty of gluon distribution and improves the fit quality up to 4.1\% from leading order to next-to-leading order and up to 1.7\% for only NLO without and with beauty and charm data contributions.  
\end{abstract}


\maketitle

\section{\label{introduction}Introduction}
 The production of heavy quarks in photoproduction ($\gamma p$) and Deep Inelastic Scattering (DIS) of ${e^\pm}p$ is one of the main tasks at HERA. For the present time the $b$-quark and the $c$-quark are the only two heavy quarks which are kInematically accessible at HERA and investigation of charm quark cross section H1-ZEUS combined data \cite{Abramowicz:1900rp} and H1 \cite{Aaron:2009af} and ZEUS \cite{Abramowicz:2014zub} beauty production data impact on the simultaneous determination of the parton distribution functions and the strong coupling, $\alpha_s(M^2_Z)$ is the main topic in this article.
 
 The charm quark mass is about $1.5$ GeV and the beauty quark mass is about $4.5$ GeV and when we compare these masses with the quantum chromodynamics scale parameter, $\Lambda_{QCD} \sim 250$~MeV we see that the masses of the heavy quarks, $m_h$ satisfy $m_h\gg\Lambda_{\rm QCD}$ and then provide a hard scale for perturbative Quantum Chromodynamics (pQCD) calculations.

 In deep inelastic ${e^\pm}p$ scattering we can approximate the ratio of photon couplings corresponding to a heavy quark, $Q_h, h=b, c$ by
\begin{equation} \label{eq:hfrac}
f(h) \sim \frac{Q_h^2}{\Sigma{Q_q^2}} 
\end{equation}
where $Q_h$ represents, only the $b$-quark and $c$-quark electric charges and $Q_q$~ with $q=u,d,s,c,b$~, represents the electric charges of kinematically accessible quark flavours. 

 Now for $b$-quark and $c$-quark we have respectively
\begin{equation} \label{eq:bfrac}
f(b) \sim \frac{Q_b^2}{Q_u^2+Q_d^2+Q_s^2+Q_c^2+Q_b^2}
        = \frac{1}{11} \simeq 0.09~, 
\end{equation}
\begin{equation} \label{eq:cfrac}
f(c) \sim \frac{Q_c^2}{Q_u^2+Q_d^2+Q_s^2+Q_c^2+Q_b^2}
        = \frac{4}{11} \simeq 0.36~.
\end{equation}

 Clearly, Eqs. \ref{eq:bfrac} and \ref{eq:cfrac} show that approximately $9$ and $36$ percent of the cross sections at HERA are originate from processes with the $b$-quark and the $c$-quark in the final state, respectively. This is our main motivation to investigate the impact of the $b$-quark and $c$-quark production data on simultaneous determination of parton distribution functions or their uncertainties and strong coupling, $\alpha_s(M^2_Z)$ in this article.

 For $f(b) \simeq 0.09~$ and $f(c) \simeq 0.36~$ to be meaningful at large enough momentum transfers, the $b$-quark and the $c$-quark  may be considered as an integral part of the so called ``quark-antiquark sea'' inside the proton, just same as the light quarks, originating from the initial state splitting of virtual gluons. On the other hand, the proton has no net beauty and charm flavour number, thus the beauty and charm quarks within the proton can only arise in pairs of $b\overline{b}$ and $c\overline{c}$. As we noted, the $c$-quark mass is about $1.5$ GeV and the b-quark mass is about $4.5$ GeV, thus at the low-energy limit the $b\overline{b}$ and $c\overline{c}$ pairs are considerably heavier than the proton mass and accordingly they cannot exist as permanent contributions within the proton.

 Of course, the consideration of the so-called intrinsic heavy quark \cite{Brodsky:1980pb}, may dramatically change changes this simple model of heavy flavor content of the proton and several theoritical and phenomenological groups \cite{Brodsky:2015uwa,Lyonnet:2015dca,Brodsky:2016tew,Brodsky:2016vig,Brodsky:2001yt,Brodsky:2000zc,Brodsky:1997fj,Vogt:1995fsa,Vogt:1995tf,Vogt:1992ki,Brodsky:2006wb,Vogt:1994zf,Dulat:2013hea,Ball:2016neh,Rottoli:2016lsg,Duan:2016rkr,Vafaee:2016jxl,Blumlein:2015qcn,Ball:2015tna,Bednyakov:2013wta,Polyakov:2015foa,Schmidt:2014gda,Bednyakov:2013zta,Hobbs:2016hpe,Brodsky:2016fyh,Koshkarev:2016acq,Laha:2016dri,Lipatov:2016feu,Boettcher:2015sqn}, currently try to reveal the probability of the existence an intrinsic heavy quark content within the proton  , but up to the development of the present analysis there is no evidence for the existence of such a contribution from HERA data \cite{Dulat:2013hea}. Therefore, in this article the beauty and charm quarks within the proton are as usual treated  as virtual quarks, which in turn arise as fluctuations from the perturbative splitting of gluons within the proton and therefore the heavy quark production may be considered as a primary probe of the gluon content of the proton.
 
 Heavy flavour PDFs play a central role in special hadronic collisions, which lead to the photons emerge from the hard parton-parton interactions in association with one ore more beauty and charm quark jets. Clearly to analyze and study these processes we need the heavy flavour PDFs. A series of experimental measurements involving beauty, charm and photon final states have recently been published by CDF and D0 Collaborations \cite{Abazov:2012ea,Abazov:2009de,D0:2012gw,Abazov:2014hoa,Aaltonen:2009wc,Aaltonen:2013ama}.

 As we noted, the $c$-quark mass is about $1.5$ GeV~ and the $b$-quark mass is about $4.5$ GeV and when we compare these values with the QCD scale value, $\Lambda_{QCD} \sim 0.25$ GeV~ we see that it is reasonable to treat the $b$-quark mass and $c$-quark mass as a hard scale in pQCD, appropriately taking into account heavy quark mass effects in perturbative calculations. Accordingly, in this article we use the full HERA run I and II combined data \cite{Abramowicz:2015mha} as a new measurements of inclusive deep inelastic scattering cross sections for our base data set and then we investigate the impact of charm quark cross section H1-ZEUS combined data \cite{Abramowicz:1900rp} and H1 \cite{Aaron:2009af} and ZEUS \cite{Abramowicz:2014zub} beauty production data on simultaneous determination of parton distribution functions and the strong coupling, $\alpha_s(M^2_Z)$. 
  
  The measurements of charm-production at HERA can be used to constrain some important parameters of QCD such as: the charm fragmentation parameters, the charm-quark mass and its running, the flavour composition of virtual quarks within the proton determination of the QCD strong coupling constant parameter, $\alpha_s(M^2_Z)$~ determination of the gluon distribution within the proton, impact on the flavour composition of quarks within the proton and their uncertainties and impact on the gluon content of the proton.
  
  The measurements of beauty-production at HERA suggest further complementary insight into the theoretical complications of heavy-flavour production in QCD. Because of higher mass of $b$-quark and smaller value of it's strong coupling constant, the pQCD behaviour of $b$-quark is slightly better than the one of $c$-quark. Over essentially the full accessible phase space at HERA, however the $b$-mass remains non-negligible and therefore a large fraction of the cross section is close to the kinematic $b\bar b$-mass threshold. Therefore, in order to obtain a reliable predictions a particularly sensitive handle on the treatment of these mass effects is required. According to Eqs. \ref{eq:bfrac} and \ref{eq:cfrac}~ the coupling of the photon to $b$-quarks is four times smaller than the coupling to $c$-quarks and the higher $b$-quark mass gives a strong kinematic suppression. Thus in practice at HERA, the $b$-quark production cross section is about $2$ times smaller than the $c$-quark production cross section.
  
 Investigation of the impact of $b$-quark and $c$-quark on the simultaneous determination of PDFs and strong coupling, $\alpha_s(M^2_Z)$ based on our methodology which is explained in section \ref{methodology} and emphasis on the central role of $\alpha_s(M^2_Z)$, on reveals these impacts, when it is considered as an extra free parameter which should be determined via QCD-fit are the main topics of this analysis.

 The outline of this  paper is as follows. In Sec.~II, we explain the theoretical framework of our analysis and discuss about the reduced cross sections and parton distribution functions. We introduce the data set used in this analysis and explain our methodology in details, in Sec.~ III. In Sec.~ IV, the impact of charm quark cross section H1-ZEUS combined data \cite{Abramowicz:1900rp} and H1 \cite{Aaron:2009af} and ZEUS \cite{Abramowicz:2014zub} beauty production data on QCD fit quality are discussed. We discus about the impact of combined  HERA I and II  sets with  and without  including charm and beauty data on simultaneous determination of the PDFs and the strong coupling, $\alpha_s(M^2_Z)$ in Sec.~V and we finally conclude with a summary in Sec.~VI.

\section{\label{dis}Cross sections and parton distributions}
Deep inelastic ${e^\pm}p$ scattering on proton at the centre-of-mass energies up to $\sqrt{s} \simeq 320\,$GeV at HERA plays a central role in pQCD. For unpolarised $e^{\pm}p$ scattering of the Neutral Current (NC) interactions, the reduced cross sections after QED radiative effects correction can be expressed in terms of the generalized structure functions by 
\begin{eqnarray}
   \sigma_{r,NC}^{{\pm}}&=&   \frac{d^2\sigma_{NC}^{e^{\pm} p}}{d{x_{\rm Bj}}dQ^2} \frac{Q^4 x_{\rm Bj}}{2\pi \alpha^2 Y_+}                                               \nonumber\\
    &=&  \tilde{F_2} \mp \frac{Y_-}{Y_+} x\tilde{F_3} -\frac{y^2}{Y_+} \tilde{F_{\rm L}}~,
    \label{eq:NC}
\end{eqnarray} 
where $\alpha$ is the fine-structure constant at zero momentum transfer, $Y_{\pm} = 1 \pm (1-y)^2$,  a helicity factor and the photon propagator are absorbed in the definitions of $\sigma_{r,NC}^{{\pm}}$~ and finally, $\tilde F_2$, $x\tilde F_3$ and $\tilde F_L$ are the generalized structure functions. Some details about these generalized structure functions have been reported in Ref~\cite{Vafaee:2017nze}.

  This article has been developed based on xFitter~\cite{xFitter} as an open source QCD framework, which is the newly updated version of the former HERAFitter package~\cite{ Sapronov}.

 Similarly, for unpolarised $e^{\pm}p$ scattering of the Charged Current (CC) interactions, the reduced cross sections after QED radiative effects correction may be expressed by
 
\begin{eqnarray}
\sigma_{r,CC}^{\pm} &=&\frac{2\pi x_{\rm Bj}}{G^2_F} \left[\frac{M^2_W+Q^2}{M^2_W}\right]^2 \frac{d^2\sigma_{CC}^{e^{\pm} p}}{d{x_{\rm Bj}}dQ^2} \nonumber \\
&= &  \frac{Y_{+}}{2}  W_2^{\pm} \mp \frac{Y_{-}}{2}x  W_3^{\pm} - \frac{y^2}{2} W_L^{\pm}~~,
\end{eqnarray}
where $\tilde W_2^{\pm}$, $\tilde W_3^{\pm}$ and $\tilde W_L^{\pm}$  are  another set of generalized structure functions and $G_F$ is the Fermi constant~\cite{Vafaee:2017nze}.

 In analogy to the inclusive NC deep inelastic $e^{\pm}p$ scattering cross section, the reduced cross sections for heavy-quark production, $\sigma_{red}^{Q\bar{Q}}$ ($Q=b,c$) may be expressed by
\begin{eqnarray}
	\sigma_{red}^{Q\bar{Q}} &=& 
	          \frac{d\sigma^{Q\bar{Q}}(e^{\pm} p)}{d{x_{\rm Bj}} \, dQ^2} \cdot \frac{x_{\rm Bj} \, Q^4}{2 \pi \alpha^2 Y_{+}} \nonumber \\ 
	          &=&F_2^{Q\bar Q} \mp \frac{Y_{-}}{Y_{+}}x  F_3^{Q\bar Q} - \frac{y^2}{Y_{+}}  F_L^{Q\bar Q}~, 
    \label{eq:NCheavy}
\end{eqnarray}	
where $\alpha$ is the electromagnetic coupling constant, $Y_{\pm} = (1 \pm (1-y)^2)$ and $F_2^{Q\bar{Q}}$, $xF_3^{Q\bar{Q}}$ and $F_L^{Q\bar{Q}}$ are heavy-quark contributions to the 
inclusive structure functions $F_2$, $xF_3$ and $F_L$, respectively.

 In the kinematic region at HERA, the $F_2^{Q\bar{Q}}$ structure function makes a dominant contribution. The $xF_3^{Q\bar{Q}}$ structure function makes the contribution only from $Z^0$ exchange and $\gamma Z^0$ and which implies for $Q^{2} \ll M_{Z}^{2}$ region, this contribution can be ignored. Finally, the contribution of longitudinal heavy-quark structure function, $F_L^{Q\bar{Q}}$~ is suppressed only for  $y^2 \ll 1$ region which may be a few percent in the kinematic region accessible at HERA and therefore cannot be ignored.
 
 Thus, neglecting the $xF_3^{Q\bar Q}$ structure function contribution the reduced heavy-quark cross section, $\sigma_{red}^{ Q\bar Q}$ for both positron and electron beams may be expressed by  
\begin{eqnarray}
   \sigma_{red}^{ Q\bar Q}&=&   \frac{d^2\sigma^{Q\bar Q}(e^{\pm}p)}{dxdQ^2} \frac{xQ^4}{2\pi\alpha^2 Y_{+}} \nonumber\\
  &= &  F_2^{Q\bar Q}  - \frac{y^2}{Y_{+}}  F_L^{Q\bar Q}~. 
    \label{eq:NCheavylast}
\end{eqnarray}
Accordingly, at high $y$ the reduced heavy-quark cross section, $\sigma_{red}^{ Q\bar Q}$, and $F_2^{Q\bar Q}$ structure function only differ by a small $F_L^{Q\bar Q}$ contribution \cite{Daum:1996ec}.

 In quark parton model~\cite{Vafaee:2017nze}, the structure functions of proton depend only on $Q^2$ variable and then may be directly related to the PDFs. Whereas, in the QCD and specially when heavy flavour productions are included the structure functions depend on both $x$ and $Q^2$ variables, \cite{Sjostrand:2001yu,Aktas:2006hy,DeRoeck:2011na,Ball:2010de,Aaron:2011gp,Tung:2006tb,Aaron:2009aa,Engelen:1998rf,CooperSarkar:2012tx,Frixione:1993yw,Marchesini:1991ch,Jung:1993gf,Sjostrand:1985ys,Sjostrand:1986hx,Lonnblad:1992tz,Kuraev:1976ge,Ciafaloni:1987ur,Jung:2000hk,Beneke:1994rs,Agashe:2014kda,Schmidt:2012az,Alekhin:2010sv,Gao:2013wwa,Martin:1998sq,Pumplin:2002vw,Chekanov:2002pv}. In section \ref{methodology}, based on our methodology we extract the PDFs based on full HERA I and II combined data, without and with the charm quark cross section H1-ZEUS combined data \cite{Abramowicz:1900rp} and H1 \cite{Aaron:2009af} and ZEUS \cite{Abramowicz:2014zub} beauty production data set included.

\section{\label{methodology}Data Set and Our Methodology}         
In this paper we use four kinds of data sets: full HERA run I and II combined NC and CC DIS $e^{\pm}p$ scattering cross sections \cite{Abramowicz:2015mha}, charm production reduced cross section measurements data \cite{Abramowicz:1900rp}, H1 \cite{Aaron:2009af} and ZEUS \cite{Abramowicz:2014zub} beauty production data sets. In our analysis we choose the full HERA run I and II combined data as our base data set, and then  we investigate the impact of charm and beauty production reduced cross section data on simultaneous determination of PDFs and the strong coupling, $\alpha_s(M^2_Z)$~.

 The combined reduced cross sections for NC and CC $e^{\pm}p$ scattering depend on the centre-of-mass energy, $\sqrt{s}$ and on the two kinematic variables $x_{\rm Bj}$ and $Q^2$ . 
The kinematic variable $x_{\rm Bj}$, in turn is related to $y$,  $Q^2$ and $s$ through the relationship $x_{\rm Bj}=Q^2/(sy)$.

 The kinematic ranges  for NC and CC $e^{\pm}p$ scattering cross sections have been reported in Ref~\cite{Vafaee:2017nze}.
 
 As we mentioned, we use the open source xFitter \cite{xFitter} version 1.2.2 as our QCD fit framework~\cite{Vafaee:2017jnt,Vafaee:2017keb}. Using the QCDNUM package \cite{Botje:2010ay}, version 17-01/12 we evolved the parton distribution functions and $\alpha_s(M^2_Z)$. In evolution of the PDFs and $\alpha_s(M^2_Z)$, we set our theory type based on DGLAP collinear evolution equations \cite{DGLAP}.
 
 In our analysis, to investigate the charm and beauty production reduced cross sections data impact we need to use the heavy flavour scheme. different theoretical groups use various heavy flavour schemes. For example, some theory groups such as CT10 \cite{Lai:2010vv}, ABKM09 \cite{Alekhin:2009vn}, NNPDF2.1 \cite{Ball:2008by,Mironov:2009uv} used  
S-ACOT \cite{Collins:1998rz}, FFNS \cite{Martin:2006qz} and  FONLL \cite{Forte:2010ta}, respectively and some other groups such as MSTW08 \cite{Martin:2009iq} and HERAPDF1.5/2.0 \cite{Aaron:2009aa} used RT standard and optimal heavy flavour schemes \cite{Thorne:2006qt,Thorne:2012az} to calculate the reduced charm and beauty cross sections in DIS.  

Also, in order to include the heavy quark flavour contribution, the perturbative QCD scales, $\mu_f^2$ and $\mu_r^2$  play a central rule. Some theoretical groups such as CT10 \cite{Lai:2010vv} and ABKM09  \cite{Alekhin:2009vn}, use  $\mu_f^2 = \mu_r^2=Q^2+m_h^2$ and $\mu_f^2 = \mu_r^2=Q^2+4m_h^2$  respectively, where $m_h$ with $h=b,c$ denotes the pole mass of $b$-quark and $c$-quark, whereas NNPDF2.1 \cite{Ball:2008by,Mironov:2009uv}, HERAPDF1.5 \cite{Aaron:2009aa} and  MSTW08 \cite{Martin:2009iq} groups use $\mu_f = \mu_r=Q$ in their heavy quark QCD approach.
  
  To include the heavy-flavor contributions, we use Thorne-Roberts Variable Flavor Number Scheme (RT-VFNS) standard and choose $\mu_f = \mu_r=Q$~ as a perturbative quantum chromodynamics scales with pole masses $m_b=4.75$ GeV and $m_c=1.5$ GeV. A full detailed study of RT variant of variable flavour number scheme may be found in Ref~\cite{Vafaee:2017nze}.  
  
   The last step in our QCD analysis is minimization of fit parameters, which we use the standard MINUIT-minimization package as a powerful program for minimization \cite{James:1975dr}.

 We parameterized the PDFs of the proton, $xf(x)$ at the initial scale of the QCD evolution $Q^2_0= 1.9$ GeV$^2$ based on the HERAPDF approach:
\begin{equation}
 xf(x) = A x^{B} (1-x)^{C} (1 + D x + E x^2)~~,
\label{eqn:pdf}
\end{equation}
where in the infinite momentum frame, $x$ is the fraction of the proton's momentum taken by the struck parton. 
  
To determine the normalization constants $A$ for the valence and gluon distributions we use the QCD number and momentum sum rules. Using this functional form, Eq. \ref{eqn:pdf} leads to the following independent combinations of parton distribution functions:
\begin{eqnarray}
\label{eq:xgpar}
xg(x) &=   & A_g x^{B_g} (1-x)^{C_g} - A_g' x^{B_g'} (1-x)^{C_g'} ,  \\
\label{eq:xuvpar}
xu_v(x) &=  & A_{u_v} x^{B_{u_v}}  (1-x)^{C_{u_v}}\left(1+E_{u_v}x^2 \right) , \\
\label{eq:xdvpar}
xd_v(x) &=  & A_{d_v} x^{B_{d_v}}  (1-x)^{C_{d_v}} , \\
\label{eq:xubarpar}
x\bar{U}(x) &=  & A_{\bar{U}} x^{B_{\bar{U}}} (1-x)^{C_{\bar{U}}}\left(1+D_{\bar{U}}x\right) , \\
\label{eq:xdbarpar}
x\bar{D}(x) &= & A_{\bar{D}} x^{B_{\bar{D}}} (1-x)^{C_{\bar{D}}}~,
\end{eqnarray}
where $xg(x)$ is the gluon distribution, $xu_{{v}}(x)$ and $xd_{{v}}(x)$ are the valence-quark distributions and $x\bar{U}(x)$ and $x\bar{D}(x)$ are the $u$-type and $d$-type anti-quark distributions which are identical to the sea-quark distributions.
A discussion about the three normalization parameters $A_{u_v},~A_{d_v}$ and $A_g$ and some additional constraints which we imposed to Eqs.~(\ref{eq:xgpar}--\ref{eq:xdbarpar}) have been reported in Ref~\cite{Vafaee:2017nze}.

\section{\label{qcdfitquality}Impact of Beauty and Charm Production Data on the QCD fit quality}
Now we investigate the impact of charm production reduced cross section measurements data \cite{Abramowicz:1900rp}, H1 \cite{Aaron:2009af} and ZEUS \cite{Abramowicz:2014zub} beauty production data sets on QCD fit quality. Really, we show that adding these data to HERA run I and II combined data~\cite{Abramowicz:2015mha} not only improves the QCD fit quality, but also reduces the error bar of gluon distributions and some of it's ratios.
To investigate the fit quality we first define $\chi^2$ function as :

\begin{equation}
\chi^2=\sum_{i=1}^{N_{\rm pts}}\left(\frac{D_i+\sum_{k=1}^{N_{\rm corr}}
r_k\sigma_{k,i}^{\rm corr}-T_i}{\sigma_i^{\rm uncorr}}\right)^2+\sum_{k=1}^{N_{\rm corr}}r_k^2,
\label{eq:chi2}
\end{equation}
where $D_i+\sum_{k=1}^{N_{\rm corr}}r_k\sigma_{k,i}^{\rm corr}$ are the data values allowed to shift by some multiple $r_k$ of the systematic error, $\sigma_{k,i}^{\rm corr}$, to give the best 
fit result, and $T_i$ are the parameterized predictions. 

As we mentioned, we use four kinds of data sets in this analysis: HERA run I and II combined data with totally 1307 data points, charm production data with totally 52 data points, H1 beauty production data with totally 12 data points and finally, ZEUS beauty production data with totally 17 data points. Now to be clear, we sometimes refer to HERA run I and II combined data as ``BASE'' and ``BASE'' plus all other remaining data sets as ``TOTAL''.

 Now the total number of data points for BASE and TOTAL are 1307 and 1388, respectively. On the other hand, we perform this QCD analysis with $Q^2 \geq {Q^2_{\rm min}=3.5}$~GeV$^2$ cut and this reduces the total number of data points from 1307 and 1388 to 1145 and 1221, respectively. Now based on Table \ref{tab:data} we can present our QCD fIt quality for leading order and next-to-leading order as follows:

for LO~:
\begin{eqnarray}
\noindent\centerline{ $\chi^2_{TOTAL}$ / dof = $\frac{1377}{1130}=1.22~~$for BASE~,} \label{lobase}\\
\noindent\centerline{ $\chi^2_{TOTAL}$ / dof = $\bf \frac{1457}{1206}= 1.21~~$for TOTAL~,} \label{lototal}
\end{eqnarray} 
and for NLO~:
\begin{eqnarray}
\noindent\centerline{ $\chi^2_{TOTAL}$ / dof = $\frac{1335}{1130}=1.18~~$for BASE~,} \label{nlobase}\\
\noindent\centerline{ $\chi^2_{TOTAL}$ / dof = $\bf \frac{1406}{1206}=1.16~~$for TOTAL~,} \label{nlototal}
\end{eqnarray}
where dof refers to the $\chi^2$ per degrees of freedom and is defined as:

dof = number of data points - number of free parameters.
As can be deduced from Eqs.~(15)-(18), we obtain up to 4.1\% improvement of $\chi^2_{TOTAL}$ / dof, from LO to NLO and up to 1.7\% only for NLO without and with charm and beauty data contributions, as our QCD fit quality. These results may be compared by a useful study which has been reported in Ref~\cite{Hou:2016nqm}, in which they used the CT14 functional parametrization of PDFs at the initial scale of $Q_0 = 1.3\, {\rm GeV}$ at both NLO and NNLO and chose the S-ACOT-chi scheme to treat the heavy-quark mass effects. They reported the values 1.22 and 1.25 for $\chi^2_{\rm HERA2}/N_{pts}$ at the NLO and NNLO, respectively in Ref~\cite{Hou:2016nqm}.

 As we will discuss in the next section this dramatic improvement of fit quality, implies a significance reduction of some PDFs uncertainties, specially for gluon distributions and some of it's ratios.

\begin{figure*}
\includegraphics[width=0.32\textwidth]{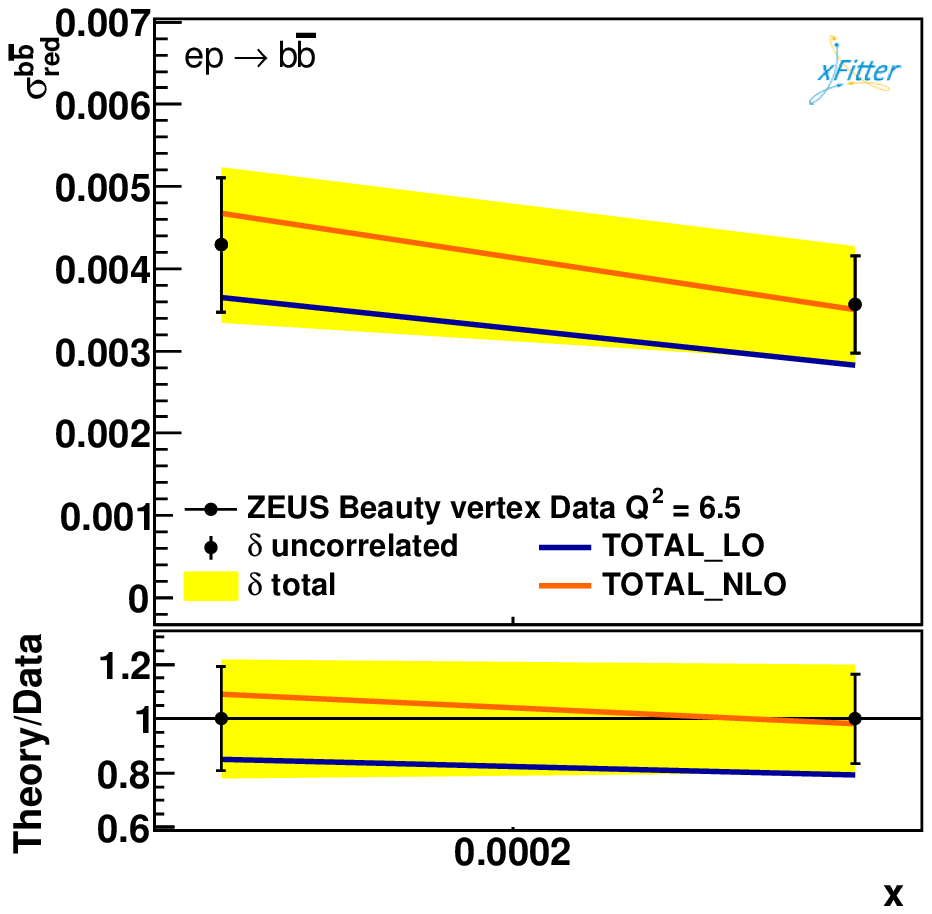}
\includegraphics[width=0.32\textwidth]{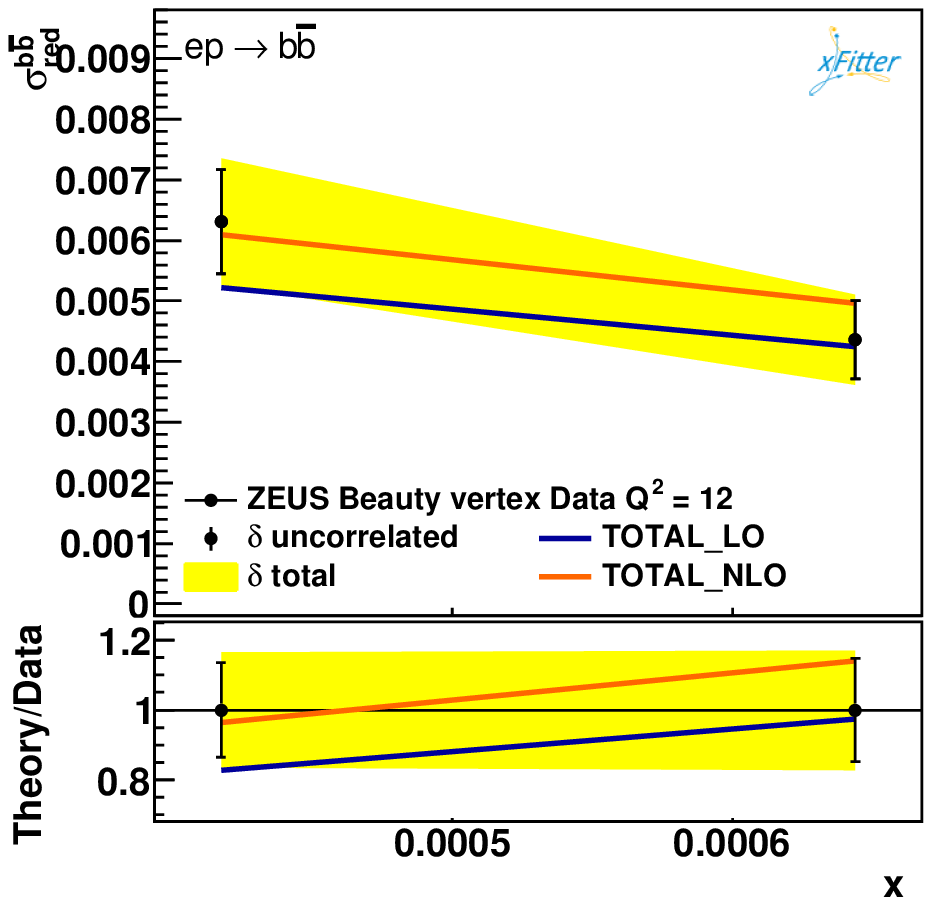}
\includegraphics[width=0.32\textwidth]{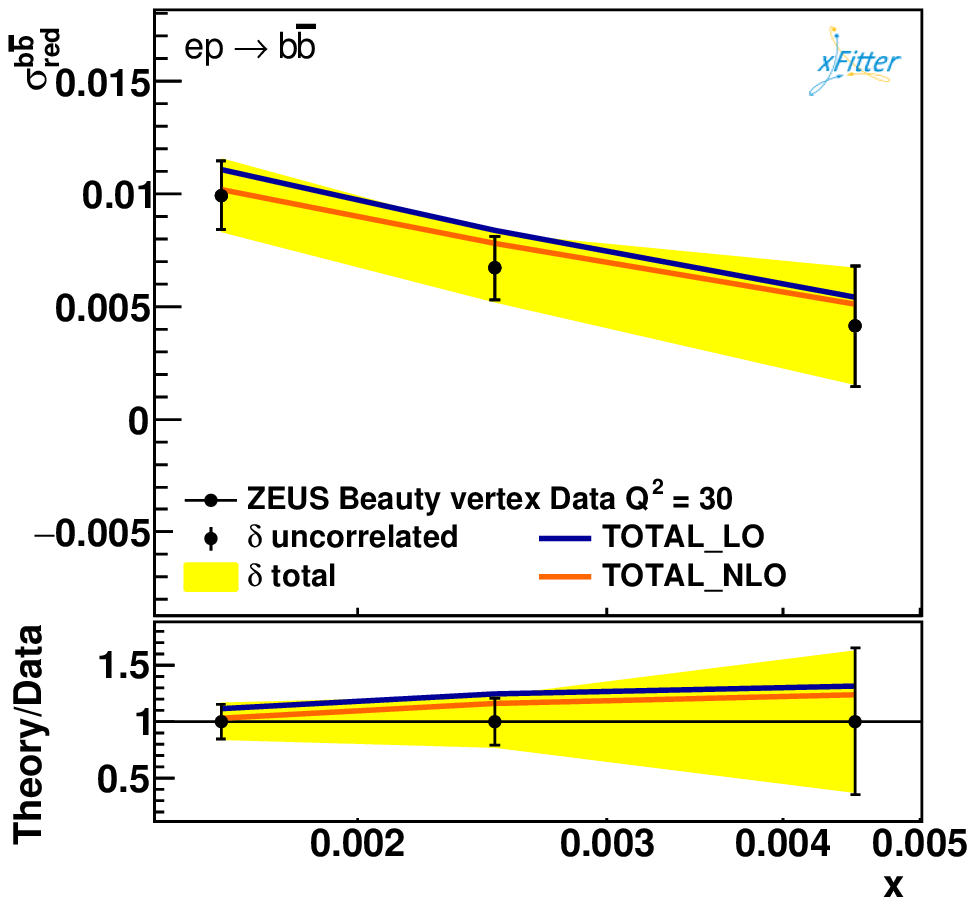}

\includegraphics[width=0.32\textwidth]{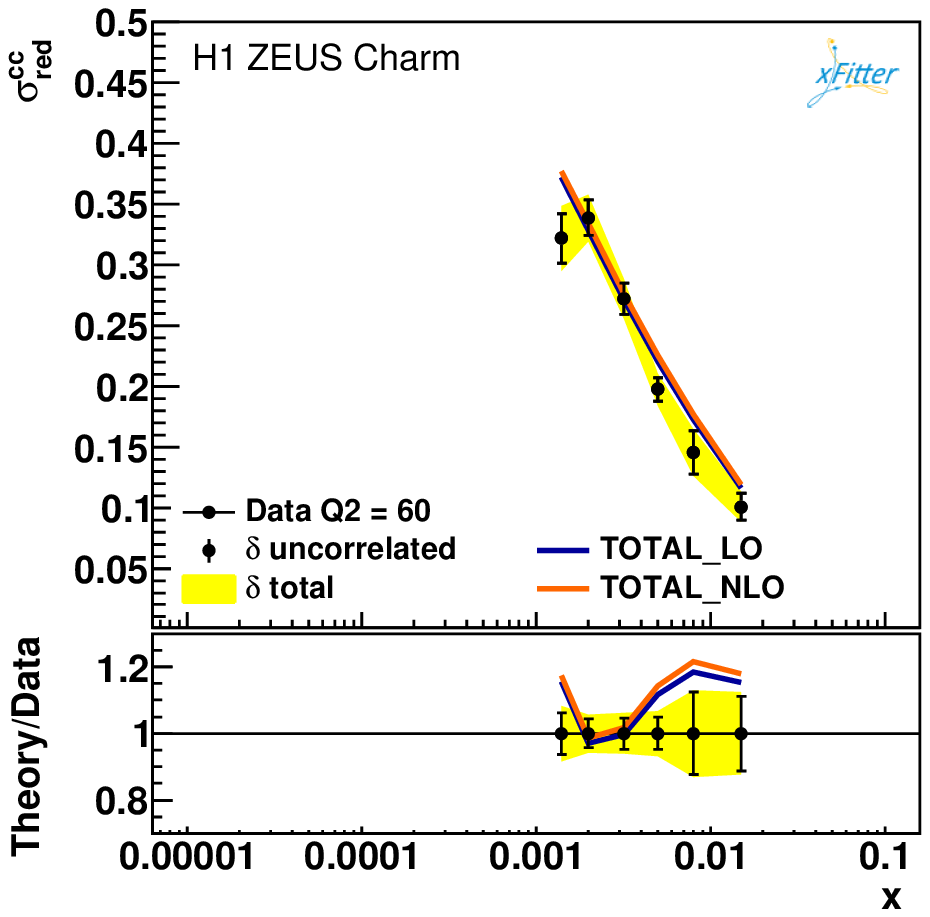}
\includegraphics[width=0.32\textwidth]{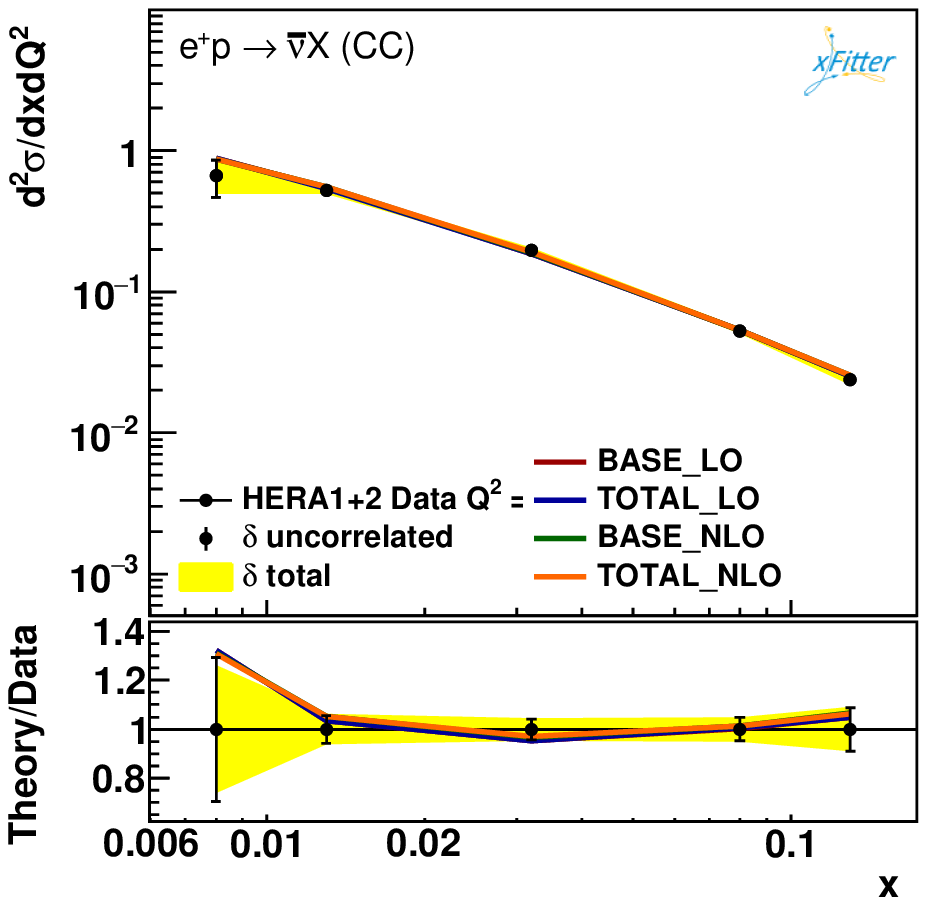}
\includegraphics[width=0.32\textwidth]{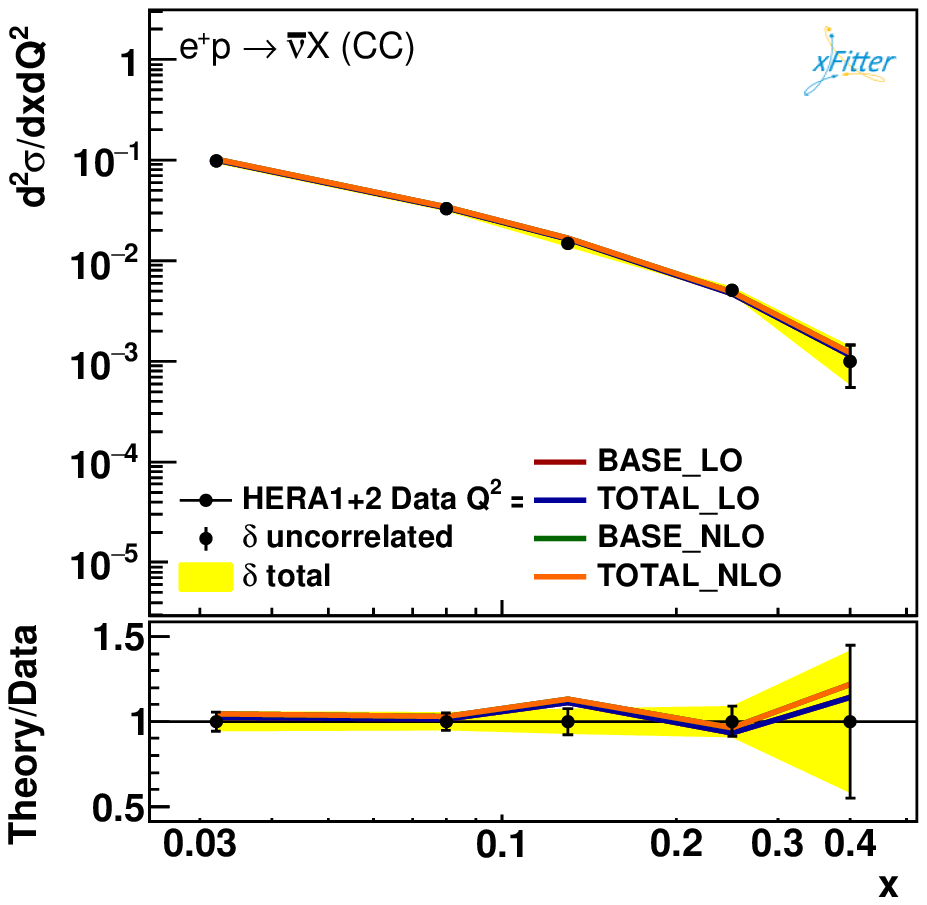}

\includegraphics[width=0.32\textwidth]{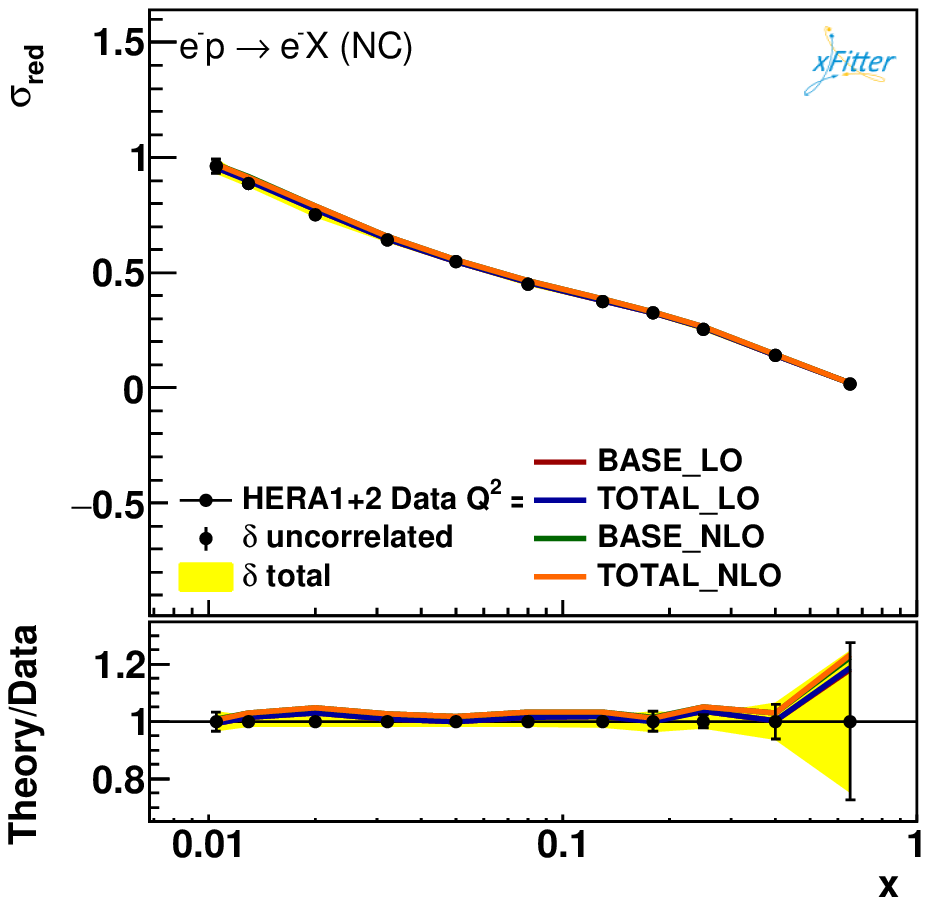}
\includegraphics[width=0.32\textwidth]{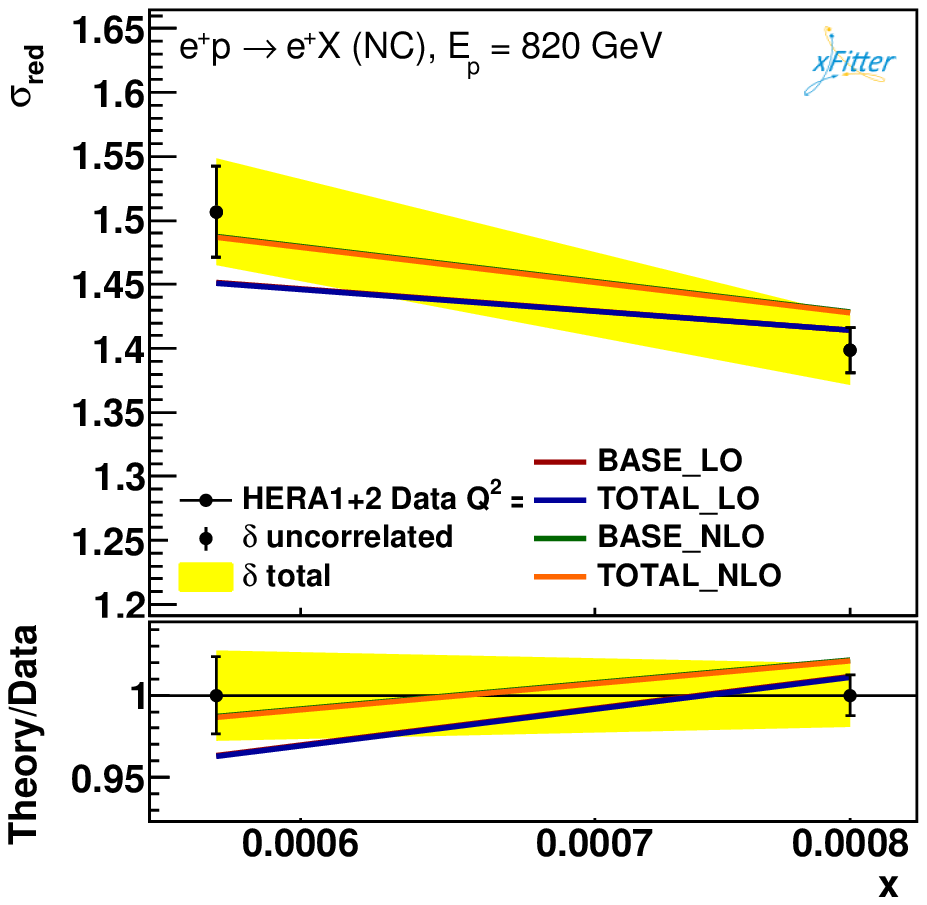}
\includegraphics[width=0.32\textwidth]{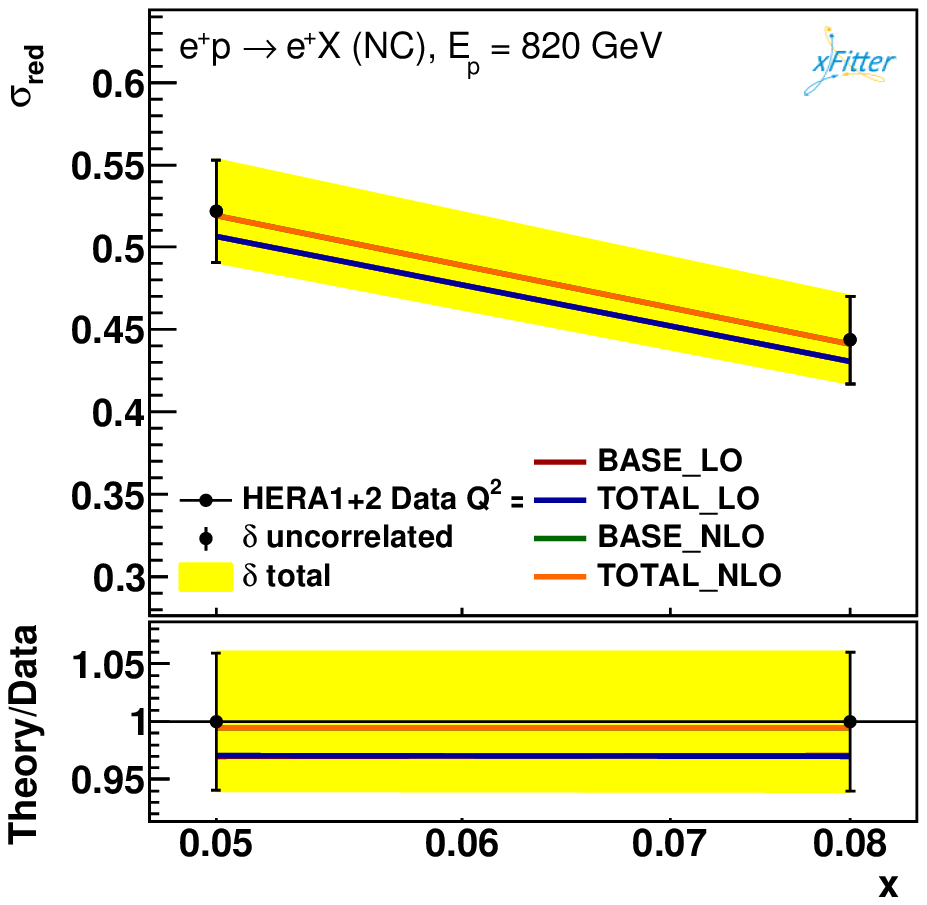}

\includegraphics[width=0.32\textwidth]{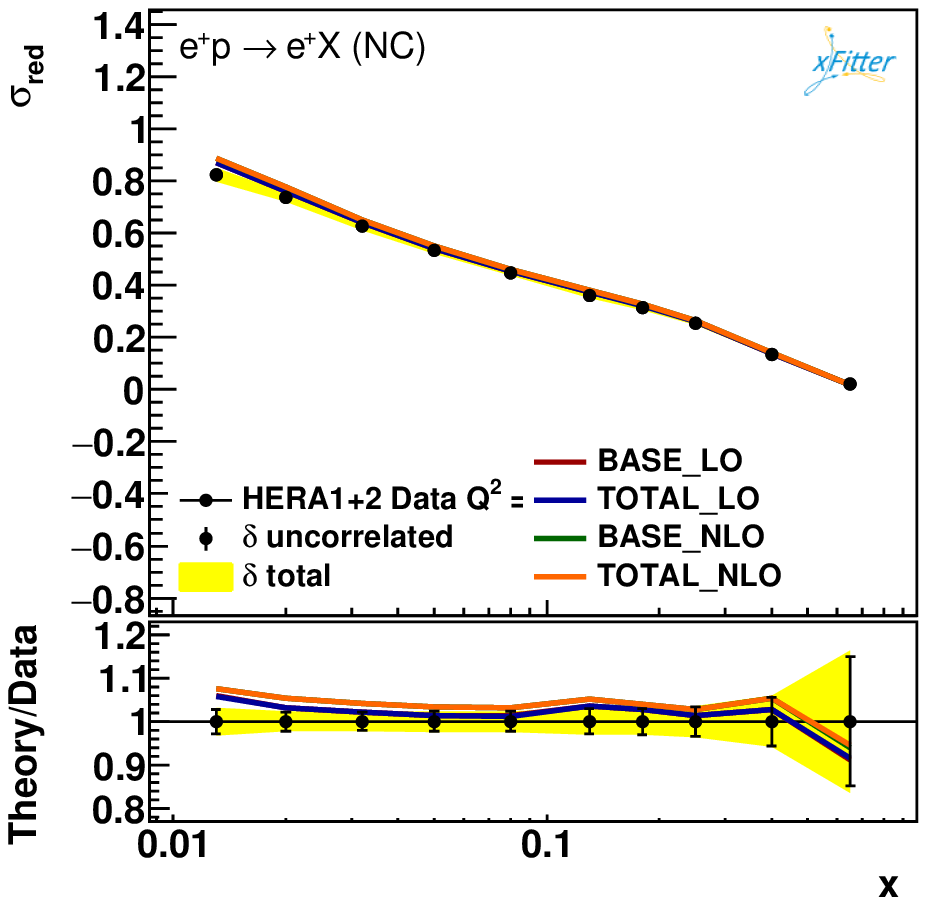}
\includegraphics[width=0.32\textwidth]{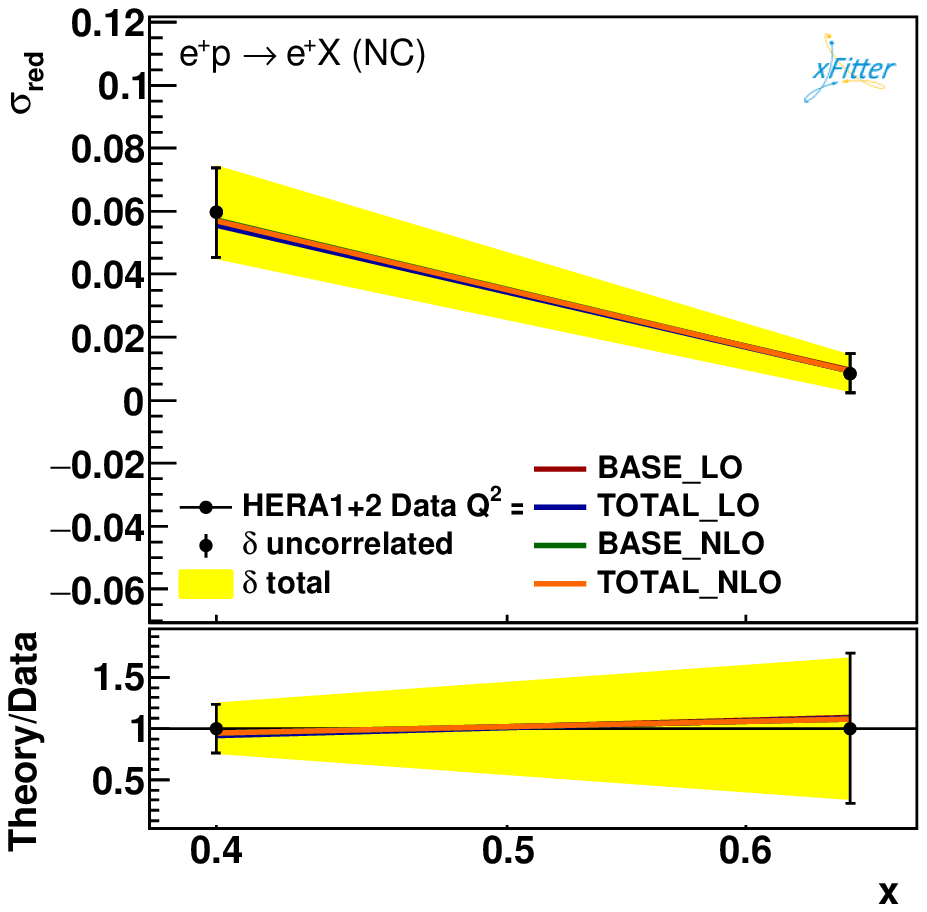}
\includegraphics[width=0.32\textwidth]{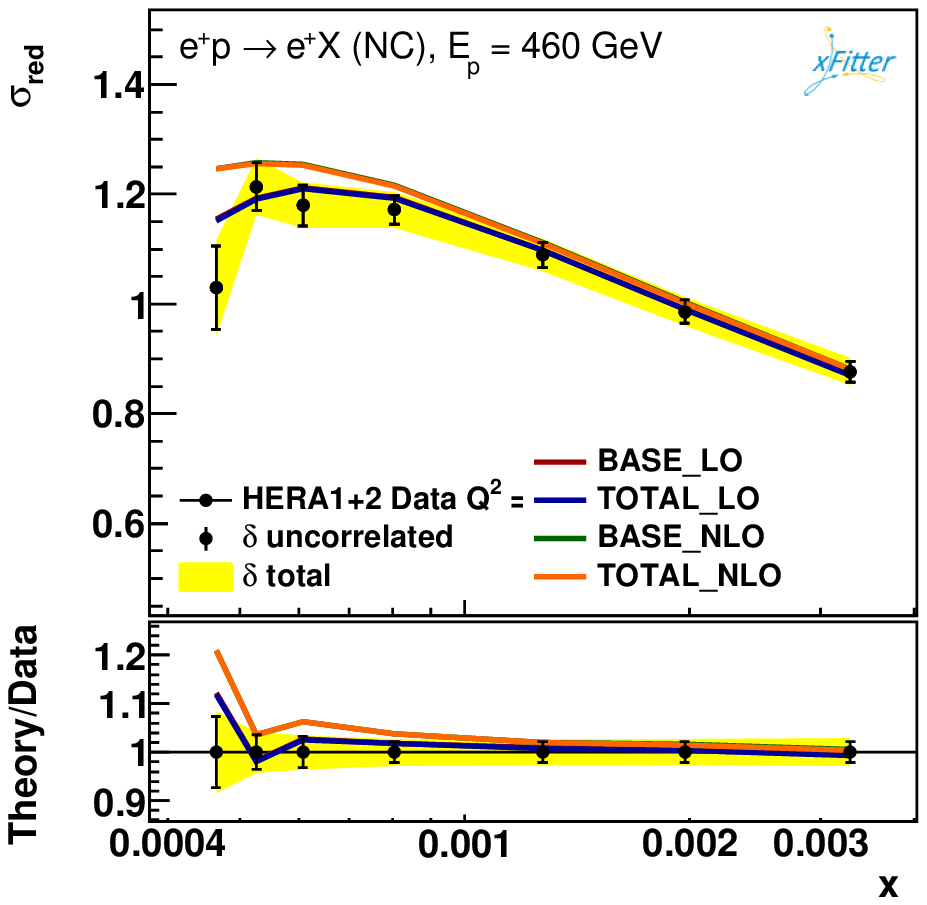}

\caption{Comparison of the  reduced deep inelastic  $e^{\pm}p$  scattering  cross sections data \cite{Abramowicz:2015mha}, charm production reduced cross section measurements data \cite{Abramowicz:1900rp}, H1 \cite{Aaron:2009af} and ZEUS \cite{Abramowicz:2014zub} beauty production data sets with our QCD fits as a function of $x$ and for different values of $Q^2$.}
\label{fig:1}
\end{figure*}

Fig. \ref{fig:1}, shows some of our QCD fit results for HERA run I and II combined inclusive DIS $e^{\pm}p$ scattering cross sections data~\cite{Abramowicz:2015mha} and charm production reduced cross section measurements data \cite{Abramowicz:1900rp}, H1 \cite{Aaron:2009af} and ZEUS \cite{Abramowicz:2014zub} beauty production data sets, as a function of $x$ and for different values of $Q^2$. These samples of QCD fit plots, show a good agreement between the theory and experiment, as we expected from our QCD fit quality.

\begin{table}[t]
\scriptsize
\caption{\label{tab:data}{Data sets used in our LO and NLO QCD analysis, with corresponding partial $\chi^2$ per data point for each data set and total $\chi^2$ per degrees of freedom (dof) for the data sets used in 15-parameter fitting.}}
\begin{ruledtabular}
\begin{tabular}{|l|c|c|c|c|c|}
 {\bf Experiment} & {\bf BASE LO} & {\bf TOTAL LO} & {\bf BASE NLO} & {\bf TOTAL NLO} \\ \hline
  HERA I+II CC $e^{+}p$ \cite{Abramowicz:2015mha} & 53 / 39& 52 / 39& 45 / 39& 45 / 39 \\ 
  HERA I+II CC $e^{-}p$ \cite{Abramowicz:2015mha} & 47 / 42& 47 / 42& 49 / 42& 49 / 42 \\ 
  HERA I+II NC $e^{-}p$ \cite{Abramowicz:2015mha} & 232 / 159& 231 / 159& 222 / 159& 221 / 159 \\ 
  HERA I+II NC $e^{+}p$ 460 \cite{Abramowicz:2015mha} & 229 / 204& 230 / 204& 209 / 204& 209 / 204 \\ 
  HERA I+II NC $e^{+}p$ 575 \cite{Abramowicz:2015mha} & 218 / 254& 218 / 254& 213 / 254& 214 / 254 \\
  HERA I+II NC $e^{+}p$ 820 \cite{Abramowicz:2015mha} & 69 / 70& 70 / 70& 66 / 70& 66 / 70 \\ 
  HERA I+II NC $e^{+}p$ 920 \cite{Abramowicz:2015mha} & 420 / 377& 422 / 377& 422 / 377& 424 / 377 \\ \hline
 {Charm H1-ZEUS} \cite{Abramowicz:1900rp} & - & 37 / 47& - & 40 / 47 \\   
{H1 beauty} \cite{Aaron:2009af} & - & 2.1 / 12& - & 2.0 / 12  \\
{{ZEUS beauty}} \cite{Abramowicz:2014zub} & - & 21 / 17& - & 11 / 17 \\ \hline
 {\bf Correlated ${\bf \chi^2}$} & 109& 127& 109& 125 \\ \hline
{\bf {Total $\bf{\chi^2}$ / dof}}  & ${\bf \frac{1377}{1130}=1.22}$ &  ${\bf \frac{1457}{1206}= 1.21}$  & ${\bf \frac{1335}{1130}=1.18}$ &  ${\bf \frac{1406}{1206}=1.16}$  \\
\end{tabular}
\end{ruledtabular}
\end{table}

\section{\label{impapdfalphs}Impact of Charm and Beauty Production Data on PDFs and the Strong Coupling}
Now we present the impact of charm production reduced cross section measurements data \cite{Abramowicz:1900rp}, H1 \cite{Aaron:2009af} and ZEUS \cite{Abramowicz:2014zub} beauty production data sets on simultaneous determination of PDFs and the strong coupling, $\alpha_s(M^2_Z)$.

 As we discussed in detail in Ref~\cite{Vafaee:2017nze}, because of correlation of the strong coupling, $\alpha_s(M^2_Z)$ with PDFs to reveal the impact of heavy flavours on parton distributions in HERAPDF approach, we perform this QCD analysis based on totally 15 free parameters, really 14 free parameters according to Eqs.~(9-13) and strong coupling, $\alpha_s(M^2_Z)$ as an extra free parameter.

In Table \ref{tab:parh}, we present the numerical values of our QCD fit of 15-free parameters and their uncertainties at the initial scale $Q^2_0 = 1.9$~GeV$^2$ and for BASE and TOTAL data sets. According to the numerical results of Table \ref{tab:parh}, when we add the charm and beauty production data sets on the  HERA run I and II combined data, the numerical value of $\alpha_s(M^2_Z)$ changes from $0.1225 \pm 0.0029$ to $0.1248 \pm 0.0027$  and from $0.1161 \pm 0.0036$ to $0.1177 \pm 0.0039$, at the LO and NLO, respectively. If we compare our main result, $\alpha_s(M^2_Z)=0.1177 \pm 0.0039$, with the world average value, $\alpha_s(M^2_Z)=0.1185 \pm 0.0006$, which has been recently reported by PDG \cite{Agashe:2014kda}, we will see that the difference is less than $\sim 0.6$\%, which is a good agreement.

\begin{figure*}
\includegraphics[width=0.23\textwidth]{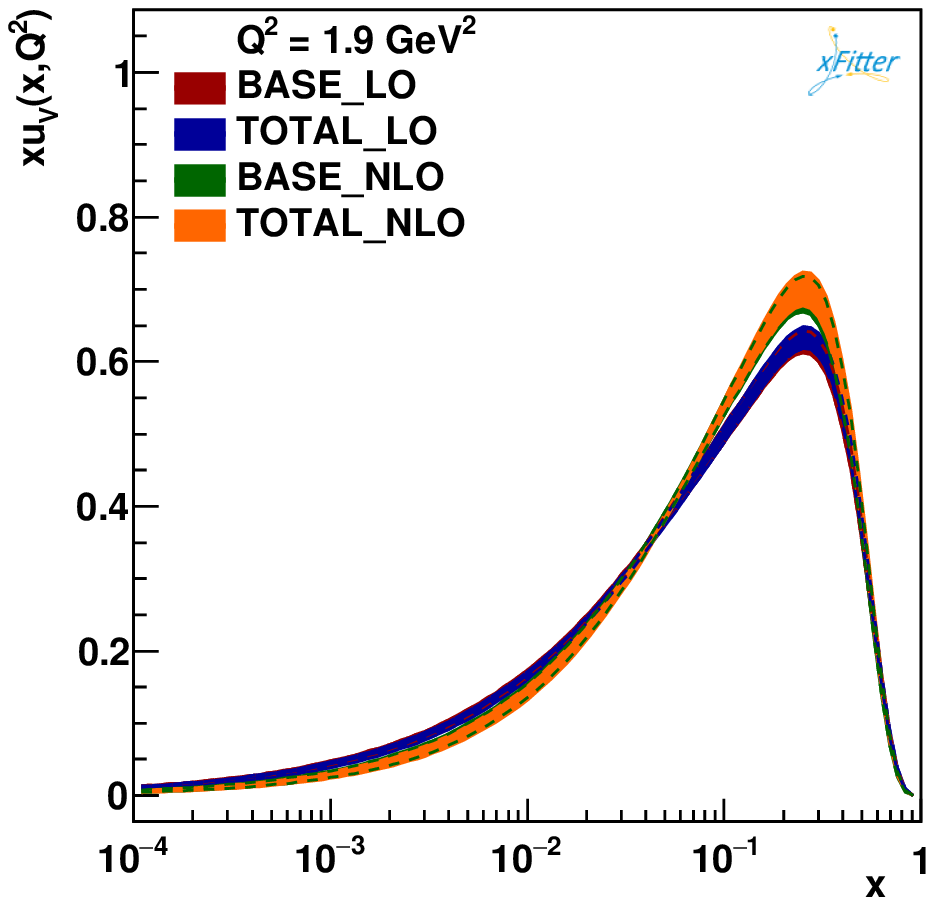}
\includegraphics[width=0.23\textwidth]{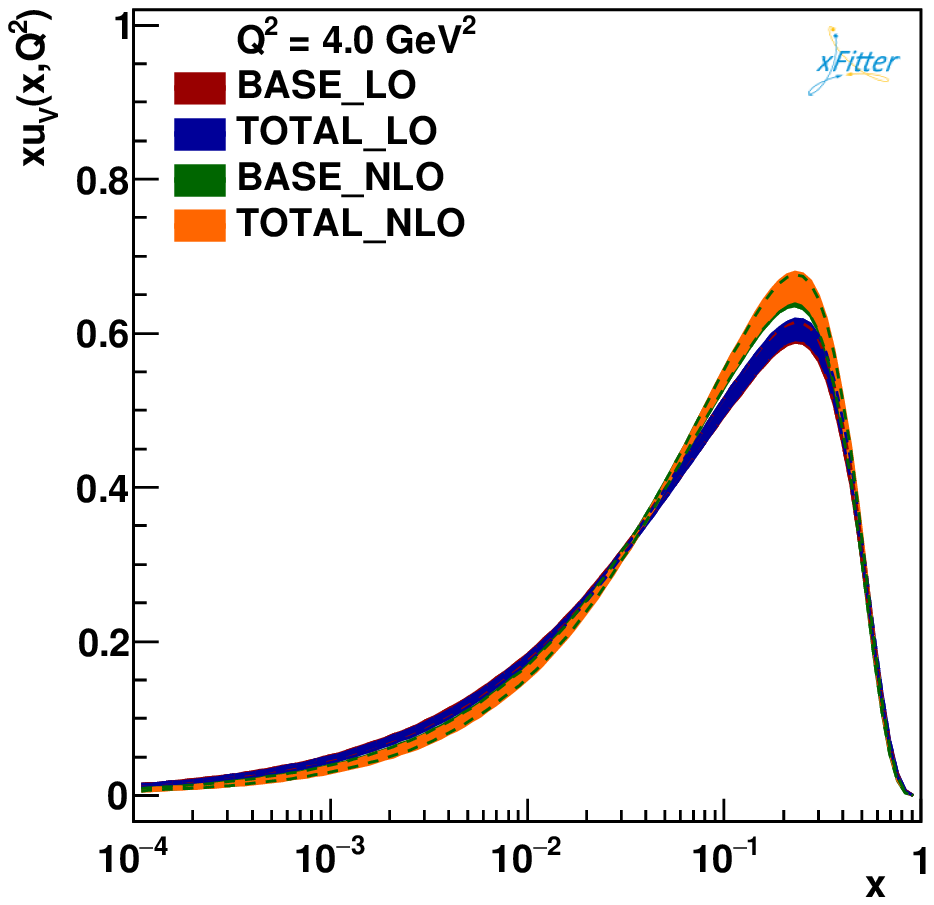}
\includegraphics[width=0.23\textwidth]{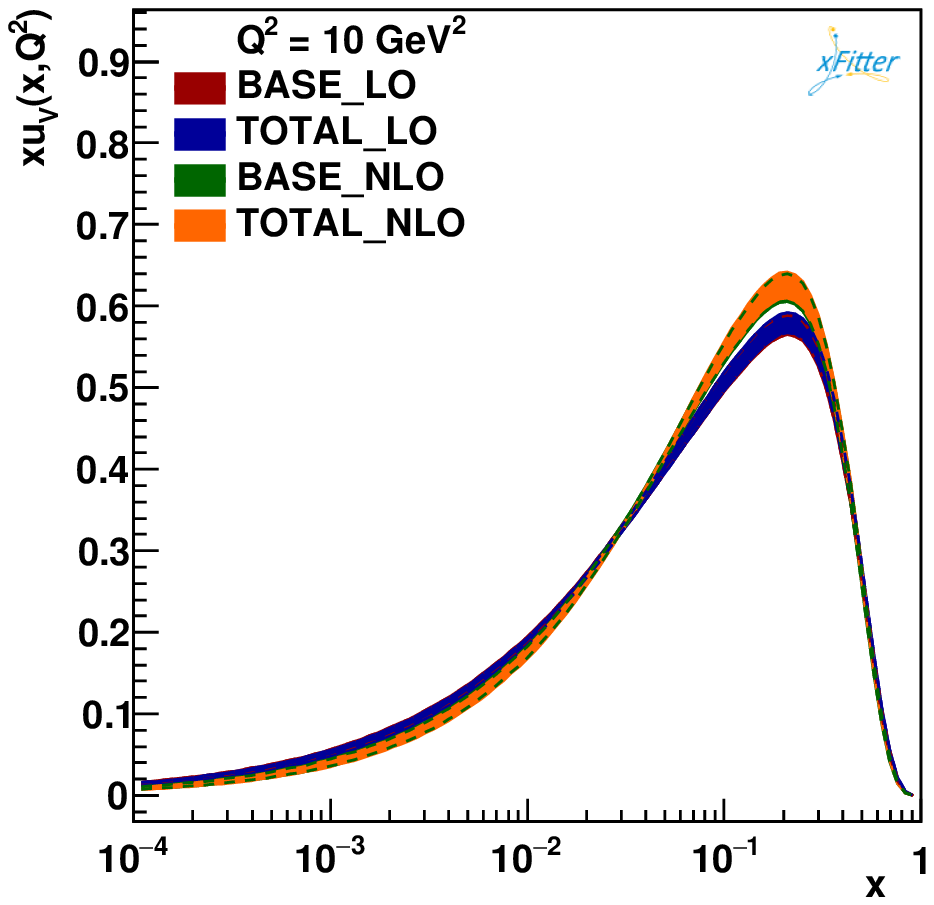}
\includegraphics[width=0.23\textwidth]{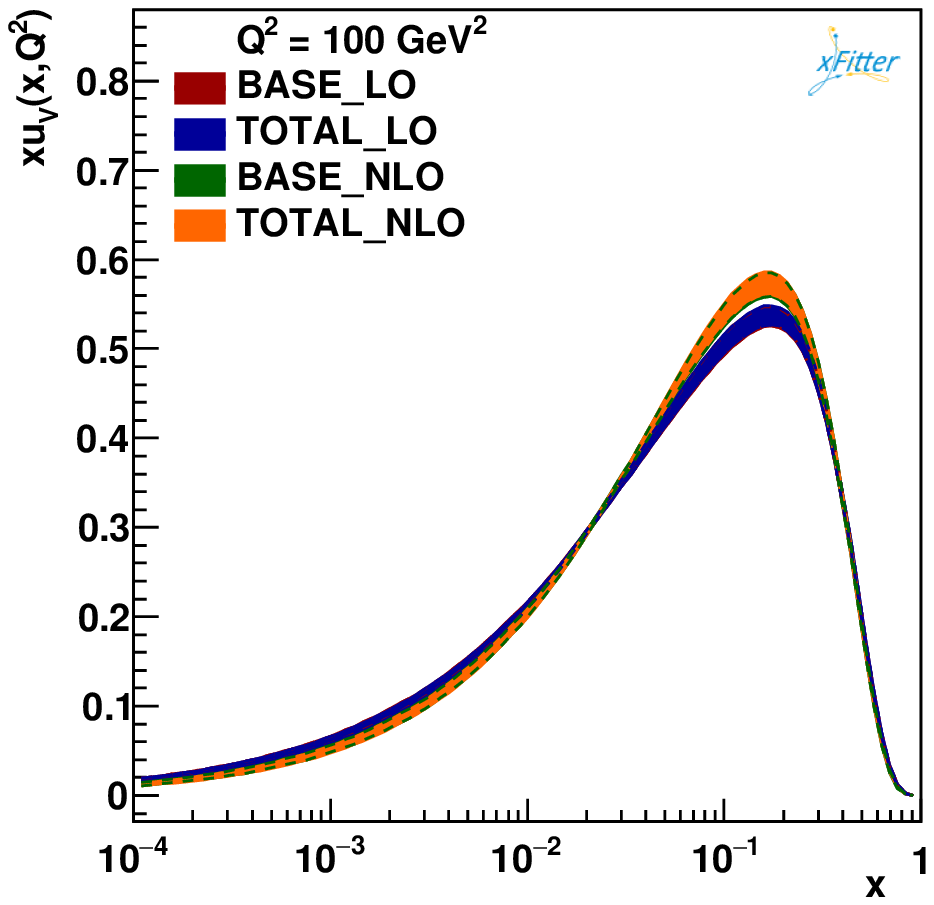}

\includegraphics[width=0.23\textwidth]{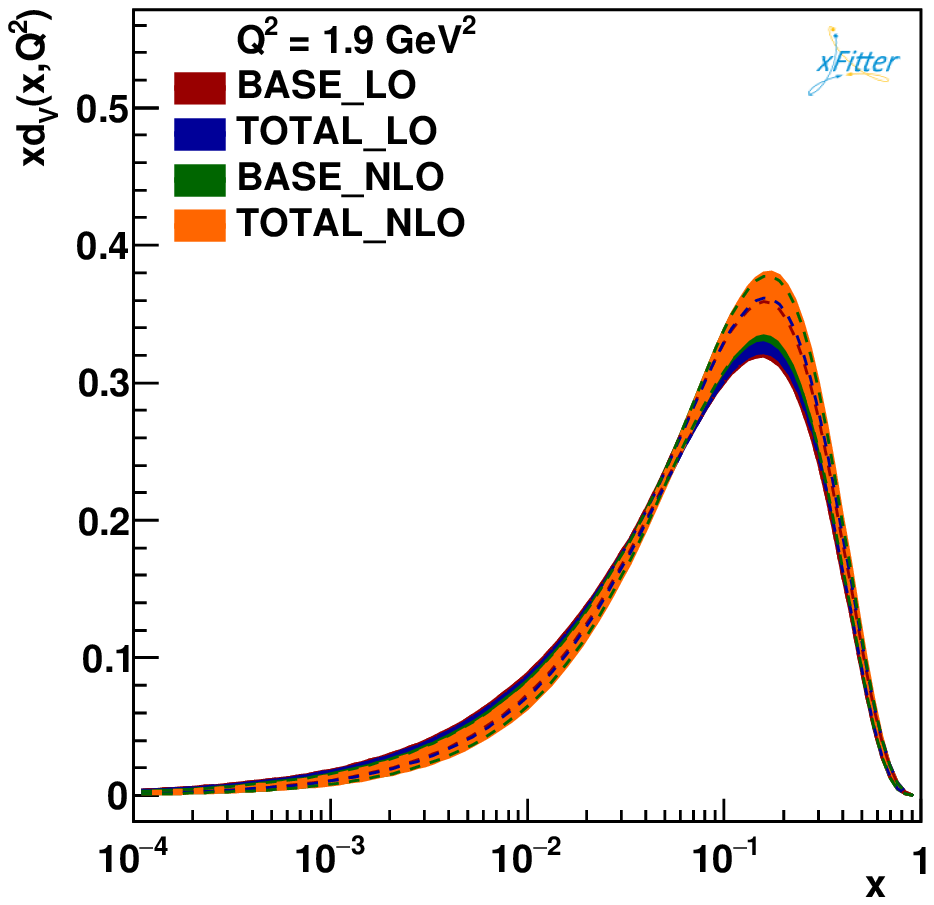}
\includegraphics[width=0.23\textwidth]{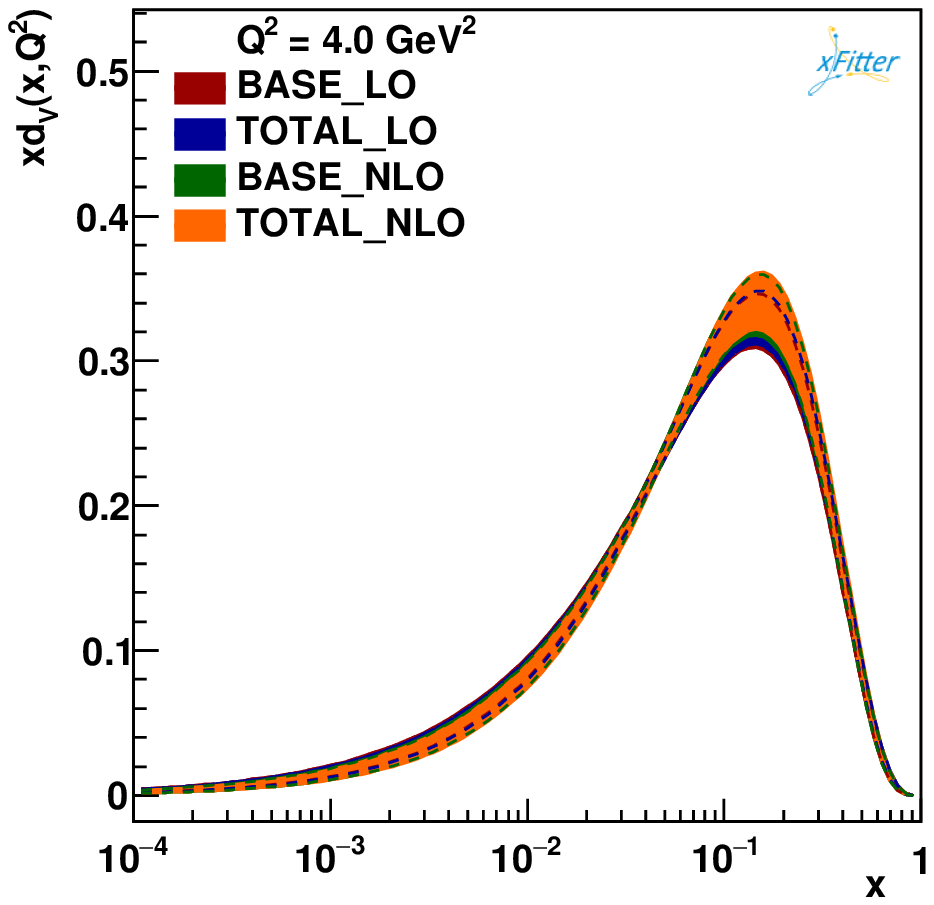}
\includegraphics[width=0.23\textwidth]{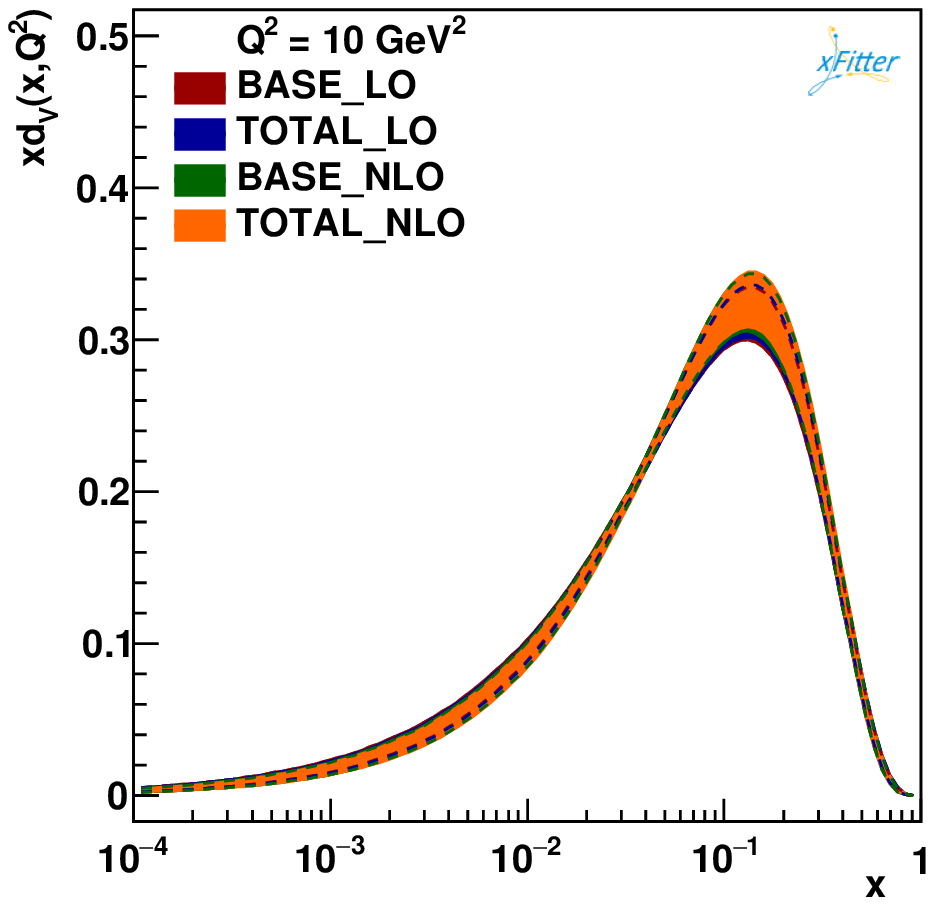}
\includegraphics[width=0.23\textwidth]{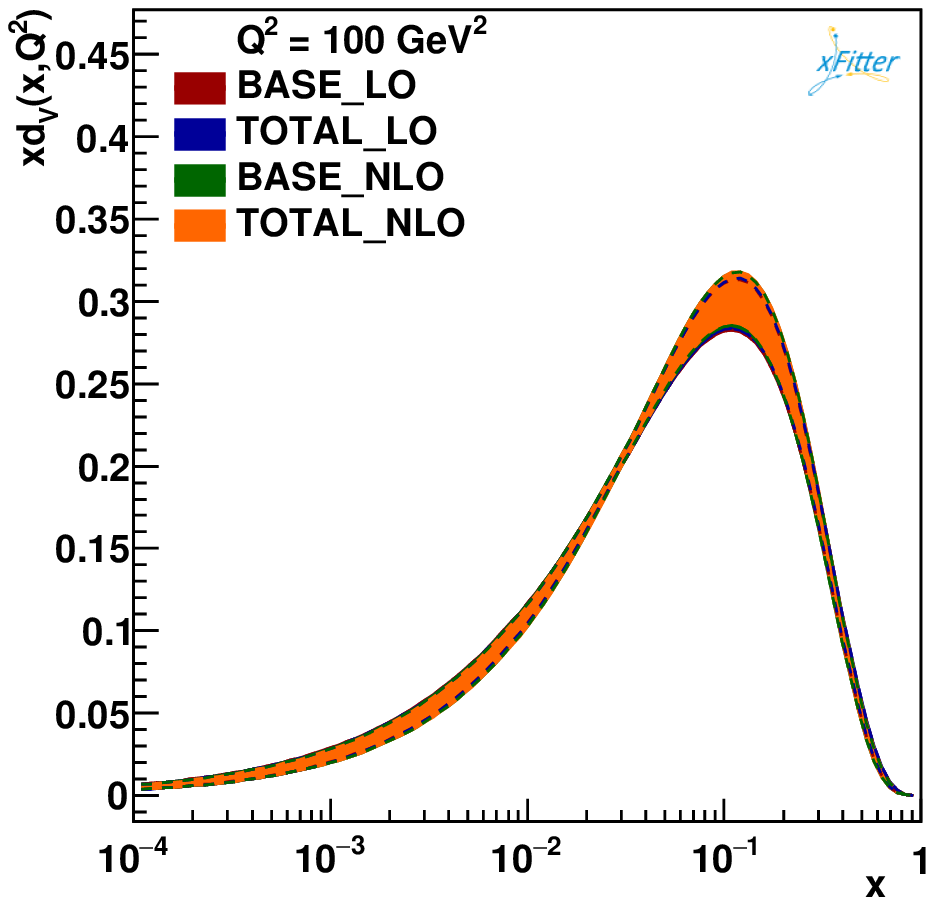}

\caption{The $xu_v$ and $xd_v$ distributions at the starting value $Q_0^2$ = 1.9~GeV$^2$ and $Q^2$ = 4, 10 and 100~GeV$^2$ as a function of $x$. The bands correspond to  PDFs  uncertainties of the fit to HERA data only (red and green) and HERA, charm and beauty data (blue and orange). As we can see, the $xu_v$ and $xd_v$ distributions are not sensitive to charm and beauty production cross sections data sets.}
\label{fig:2}
\end{figure*}

In Fig. \ref{fig:2}, we show $xu_v$ and $xd_v$ distributions at the initial scale $Q_0^2$ = 1.9~GeV$^2$ and $Q^2$ = 4, 10 and 100~GeV$^2$, as a function of $x$ at the leading order and next-to-leading order. According to the numerical results in Table \ref{tab:parh}, when we compare the numerical values of PDFs parameters for $xu_v$ and $xd_v$ at the LO and NLO, without and with charm and beauty data sets included, we do not expect to see a remarkable impact of heavy flavour contributions data on $u-$ valence and $d-$ valence quarks distributions.  Accordingly, as we see in Fig \ref{fig:2}, the $xu_v$ and $xd_v$ distributions are not sensitive to charm and beauty measurements data.  

\begin{figure*}
\includegraphics[width=0.32\textwidth]{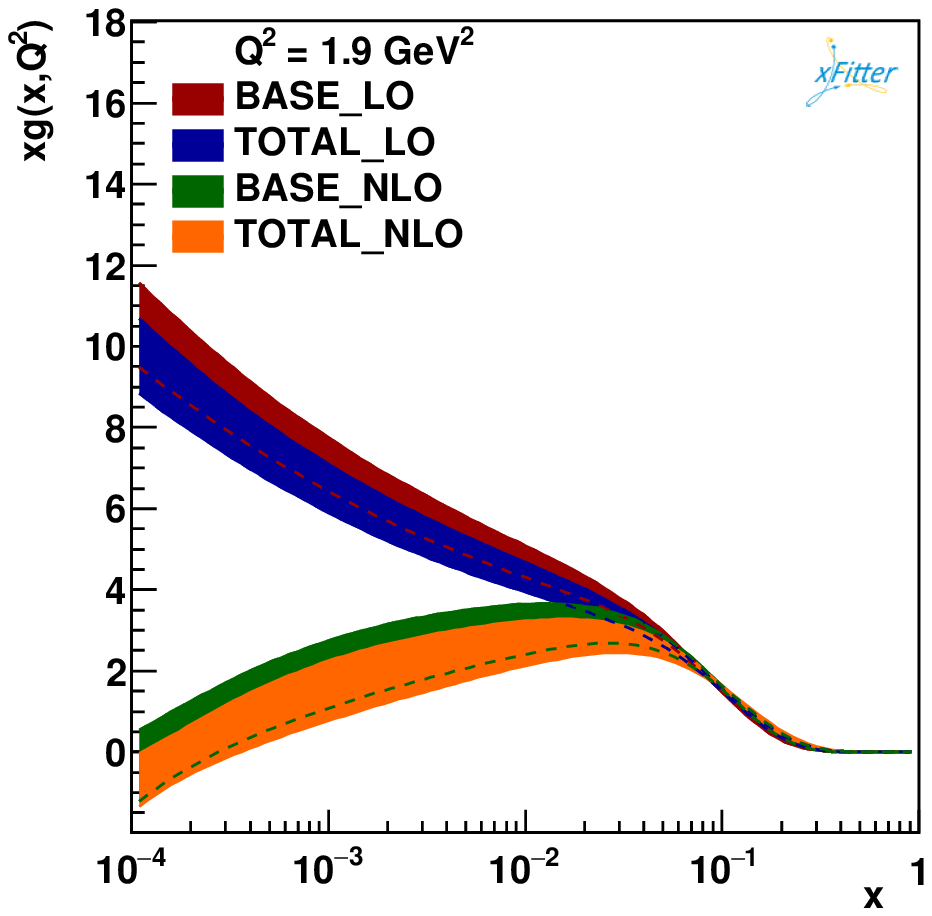}
\includegraphics[width=0.32\textwidth]{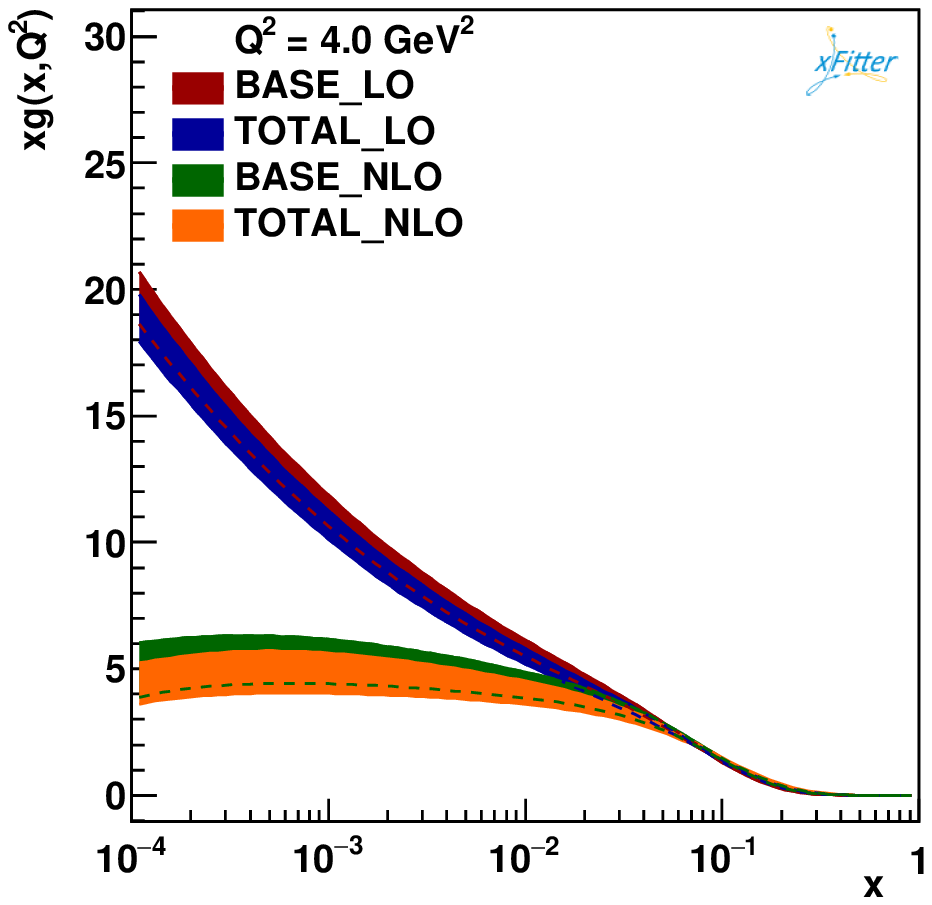}
\includegraphics[width=0.32\textwidth]{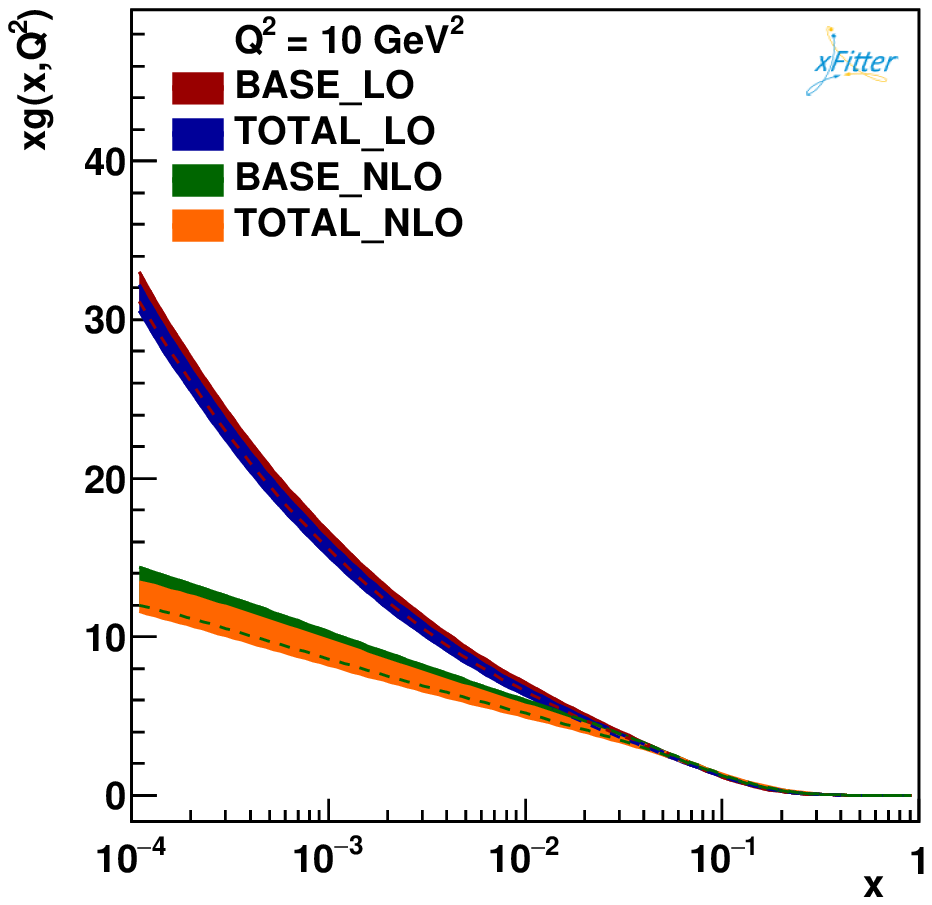}
\caption{The gluon PDFs at the $Q^2$ = 1.9, 4 and 10~GeV$^2$ as a function of $x$ at LO and NLO. The bands correspond to gluon PDF uncertainties of the fit to HERA data only (red and green) and HERA, charm and beauty data sets (blue and orange).}
\label{fig:3}
\end{figure*}

The impact of charm and beauty production cross sections data sets on HERA I and II combined data, for gluon distribution functions ($xg$) are shown in Fig. \ref{fig:3}, and for $Q^2$ = 1.9, 4 and 10~GeV$^2$ at the leading order and next-to-leading order. As we expected from our numerical values of QCD fit results for gluon parameters in Table \ref{tab:parh}, the gluon distribution functions are sensitive to charm and beauty data sets and as we see in Fig. \ref{fig:3}, adding the heavy flavour contribution data sets, dramatically reduces the uncertainty band in comparison to this band, when only HERA I and II combined data are included. 

\begin{figure*}
\includegraphics[width=0.32\textwidth]{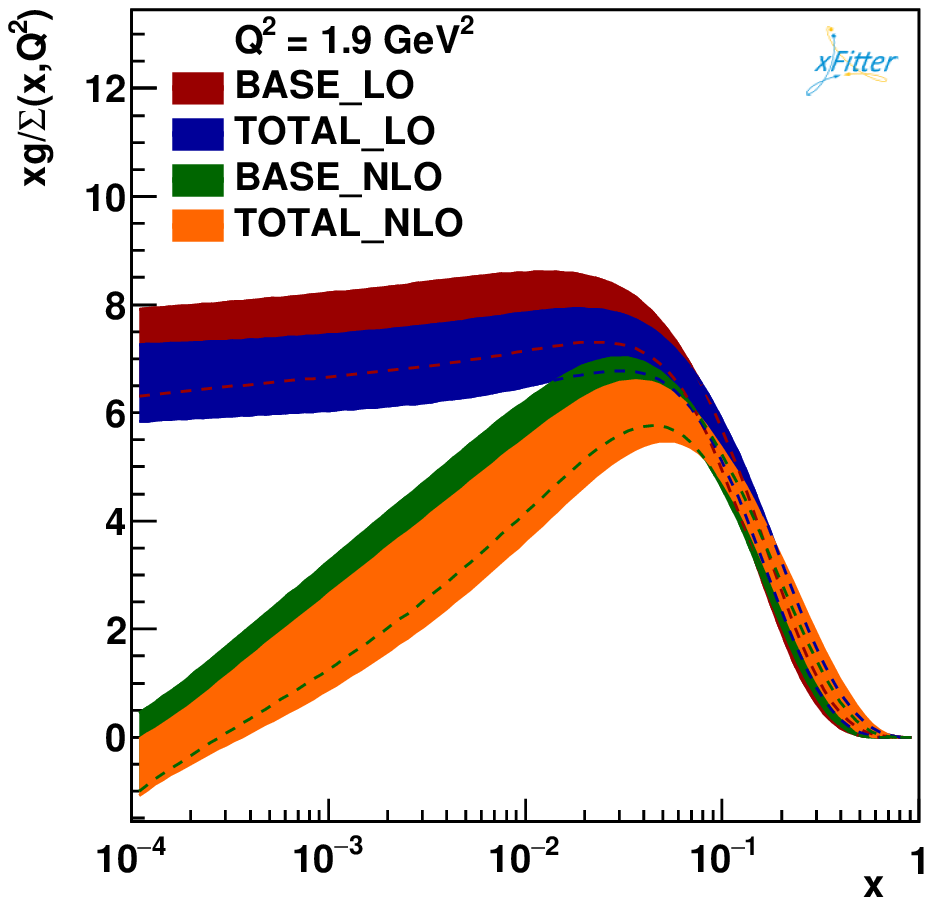}
\includegraphics[width=0.32\textwidth]{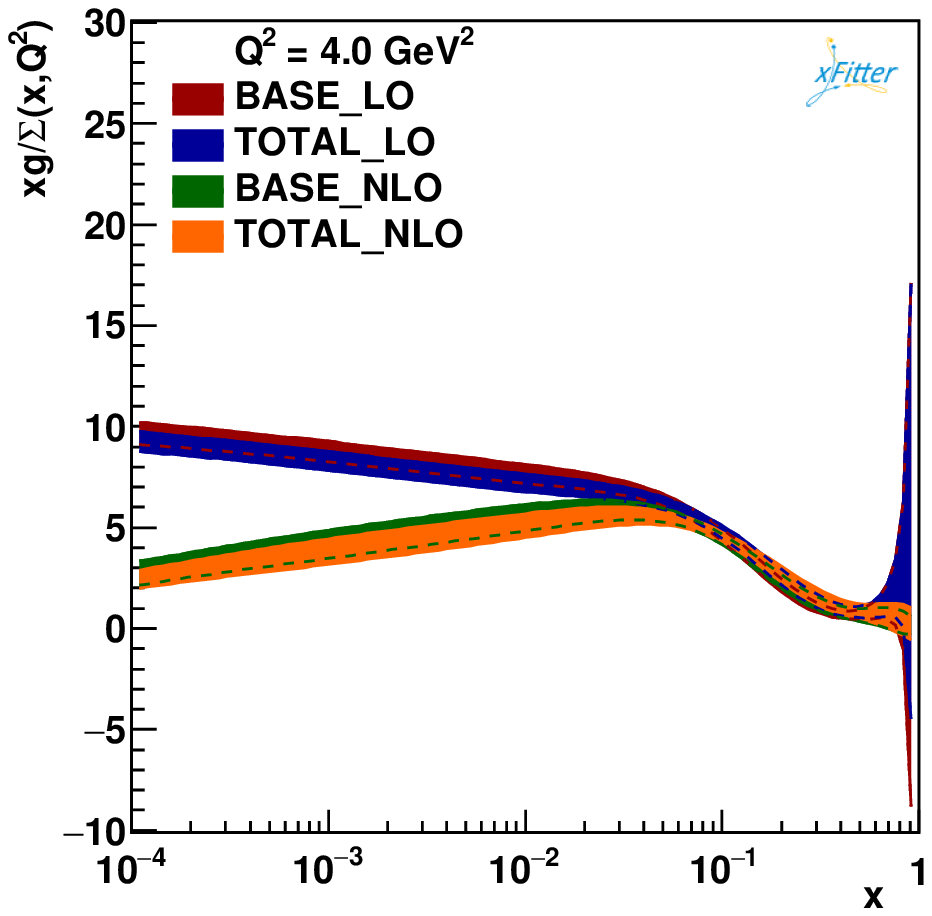}
\includegraphics[width=0.32\textwidth]{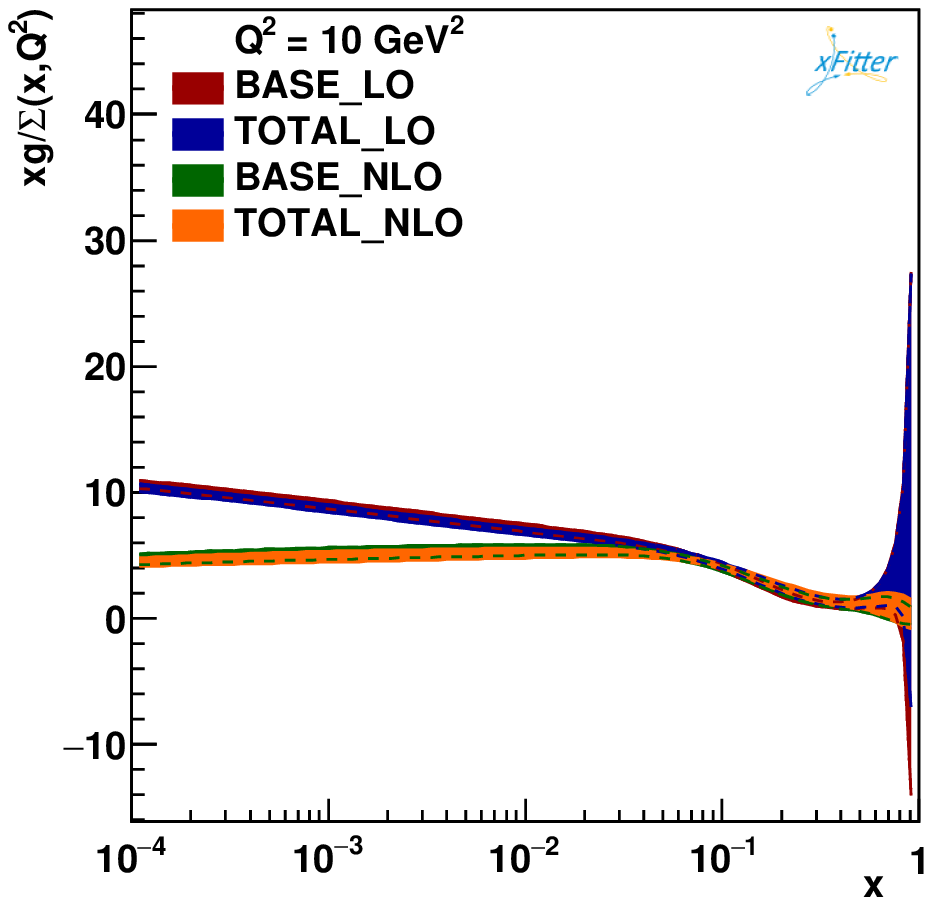}
\caption{The ratio of $xg$ (gluon distribution) over $\Sigma$ (sea quark) PDFs, without and with the charm and beauty data sets included at the $Q^2$ = 1.9, 4 and 10~GeV$^2$ as a function of $x$, at LO and NLO. The bands correspond to ratio of $xg$ over $\Sigma$ PDF uncertainties of the fit to HERA data only (red and green) and HERA, charm and beauty data sets (blue and orange).}
\label{fig:4}
\end{figure*}

The sea quark $\Sigma$-PDFs, is defined by $\Sigma=2x(\bar u+\bar d+\bar s+\bar c)$ and in Fig. \ref{fig:4}, we show the ratio of $xg$ (gluon distribution) over $\Sigma$-PDFs, without and with the charm and beauty production data sets included for $Q^2$ = 1.9, 4 and 10~GeV$^2$ at leading order and next-to-leading order. As an Impact of heavy flavour contribution data sets on these ratio, we see that the error band of uncertainties reduces considerably both at LO and NLO.

\begin{table}[t]
\scriptsize
\caption{\label{tab:parh}{ {The LO and NLO numerical values of parameters and their uncertainties for  the 
$xu_v$, $xd_v$, $x\bar u$, $x\bar d$, $x\bar s$ and $xg$ PDFs at the initial scale 0f $Q^2_0 = 1.9$~GeV$^2$.}}}
\begin{ruledtabular}
\begin{tabular}{|l|c|c|c|c|c|}
 {\bf Parameter} & {\bf BASE LO} & {\bf TOTAL LO} & {\bf BASE NLO} & {\bf TOTAL NLO} \\ \hline  
  ${B_{u_v}}$ & $0.605 \pm 0.032$& $0.609 \pm 0.033$& $0.712 \pm 0.047$& $0.723 \pm 0.046$ \\ 
  ${C_{u_v}}$ & $4.572 \pm 0.096$& $4.529 \pm 0.093$& $4.88 \pm 0.12$& $4.83 \pm 0.11$ \\ 
  $E_{u_v}$ & $14.4 \pm 2.2$& $14.3 \pm 2.2$& $13.9 \pm 2.3$& $13.5 \pm 2.6$ \\ 
  ${B_{d_v}}$ & $0.757 \pm 0.080$& $0.765 \pm 0.082$& $0.811 \pm 0.094$& $0.824 \pm 0.094$ \\ 
  $C_{d_v}$ & $4.07 \pm 0.36$& $4.04 \pm 0.37$& $4.18 \pm 0.38$& $4.17 \pm 0.42$  \\ 
  $C_{\bar{u}}$ & $7.73 \pm 0.72$& $7.33 \pm 0.72$& $9.09 \pm 0.89$& $8.67 \pm 0.95$ \\ 
  $D_{\bar{u}}$ & $13.6 \pm 3.2$& $11.4 \pm 2.9$& $18.5 \pm 3.9$& $16.2 \pm 3.6$  \\ 
  $A_{\bar{D}}$ & $0.1434 \pm 0.0079$& $0.1470 \pm 0.0080$& $0.160 \pm 0.012$& $0.161 \pm 0.013$  \\ 
  $B_{\bar{D}}$ & $-0.1984 \pm 0.0061$& $-0.1966 \pm 0.0060$& $-0.1657 \pm 0.0099$& $-0.166 \pm 0.012$  \\ 
  $C_{\bar{D}}$ & $5.8 \pm 1.4$& $6.0 \pm 1.5$& $4.4 \pm 1.4$& $4.5 \pm 1.3$ \\
  $B_g$ & $-0.171 \pm 0.046$& $-0.191 \pm 0.048$& $-0.13 \pm 0.11$& $-0.12 \pm 0.20$  \\ 
  $C_g$ & $13.0 \pm 2.0$& $11.3 \pm 1.7$& $11.8 \pm 2.2$& $10.5 \pm 2.7$  \\ 
  $A_g'$ & $2.8 \pm 1.5$& $1.95 \pm 0.91$& $2.2 \pm 1.1$& $1.8 \pm 1.6$  \\ 
  ${B_g'}$ &  $-0.164 \pm 0.078$& $-0.195 \pm 0.080$& $-0.217 \pm 0.074$& $-0.217 \pm 0.095$  \\  
  \hline 
  {${\bf \alpha_s(M^2_Z)}$} & ${\bf 0.1225 \pm 0.0029}$& ${\bf 0.1248 \pm 0.0027}$ & ${\bf 0.1161 \pm 0.0036}$& ${\bf 0.1177 \pm 0.0039}$ \\
  \end{tabular}
\end{ruledtabular}
\end{table}

\section{\label{Summary}Summary}
Up to $9$ and $36$ percent of the cross sections at HERA are originate from processes with beauty and charm quarks, respectively, in the final state. In this QCD analysis we investigate the impact of charm and beauty production cross sections on the PDFs and determination of the strong coupling, $\alpha_s(M^2_Z)$.

The sensitivity of PDFs uncertainties to charm production reduced cross section measurements data, H1 and ZEUS beauty production data sets at leading order and next-to-leading order, specially when we take the strong coupling, $\alpha_s(M^2_Z)$ as an extra free parameter is reported in this QCD analysis.

 Our extracted next-to-leading order PDFs at the initial scale, $Q^2_0$ indicate the greater sensitivity of gluon, charm and beauty PDFs uncertainty than n the case of leading-order. However, the shapes of LO and NLO PDFs of both main and without charm and beauty quarks cross sections data agree reasonably well.

 This analysis shows a dramatic reduction of some PDFs uncertainty and improvement of fit quality up to 4.1\% from LO to NLO and up to 1.7\% only for NLO without and with the charm and beauty data contributions. Also we show that reduced charm and beauty quarks cross sections can improve the uncertainty of the ratio of fitted gluon and some distributions extracted using the combined all data sets and  HERA run I and II  data only.

 A standard LHAPDF library file of this QCD analysis at the leading order and next-to-leading order is available and can be obtained via e-mail from the authors.
 


\clearpage

\end{document}